\documentclass[10pt,journal,compsoc]{IEEEtran}
\usepackage{algpseudocode}

\usepackage{graphicx}

\usepackage{float}
\usepackage{array}
\usepackage{tabularx}
\usepackage{balance}
\usepackage{marvosym}
\usepackage{amssymb}
\usepackage{amsmath}
\usepackage{multirow}
\usepackage{array}
\usepackage{enumitem}
\usepackage{booktabs} % For formal tables
\usepackage[ruled,vlined]{algorithm2e}
\usepackage{amsmath,amssymb,amsfonts}
\usepackage{graphicx}
\usepackage{subfigure}
\usepackage{textcomp}
\usepackage{xcolor}
\usepackage{pifont}
\usepackage{url}
\usepackage{bm}
\usepackage{makecell}
\usepackage{caption}
\makeatletter
\newcommand{\rmnum}[1]{\romannumeral #1}
\newcommand{\Rmnum}[1]{\expandafter\@slowromancap\romannumeral #1@}
\makeatother

\begin{document}
	
	\title{Self-Supervised Learning for Recommender Systems: A Survey}
	
	\author{Junliang~Yu, Hongzhi~Yin$^{*}$, Xin Xia, Tong Chen, Jundong Li, and Zi Huang
		\IEEEcompsocitemizethanks{\IEEEcompsocthanksitem J. Yu, H. Yin, X. Xia, T. Chen, and Zi Huang are with the School of Information Technology and Electrical Engineering, The University of Queensland, Brisbane, Queensland, Australia.\protect\\
			E-mail: \{jl.yu, h.yin1, x.xia, tong.chen, helen.huang\}@uq.edu.au
			\IEEEcompsocthanksitem J. Li is with Department of Electrical and Computer Engineering, Department of Computer Science, and School of Data Science, University of Virginia, Charlottesville, Virginia, US.\protect\\
			E-mail: jundong@virginia.edu					
		}% <-this % stops an unwanted space
		\thanks{$^{*}$Corresponding author.}
	}

	% The paper headers
	\markboth{IEEE TRANSACTIONS ON KNOWLEDGE AND DATA ENGINEERING}%
	{Shell \MakeLowercase{\textit{et al.}}: Bare Demo of IEEEtran.cls for Computer Society Journals}

	\IEEEtitleabstractindextext{%
		\begin{abstract} 
			% The neural architecture-based recommenders have achieved tremendous success in recent years and demonstrate overwhelming advantages over their conventional counterparts. However, when dealing with highly sparse data, deep recommendation models often fail to reach their full potential. Recently, self-supervised learning (SSL), emerging as a new paradigm that can automatically discover supervisory signals from insufficient raw data with well-designed pretext tasks, has become the latest trend in multiple fields. As the feature of SSL is well-aligned with recommender systems' needs for more training data, 
			In recent years, neural architecture-based recommender systems have achieved tremendous success, but they still fall short of expectation when dealing with highly sparse data. Self-supervised learning (SSL), as an emerging technique for learning from unlabeled data, has attracted considerable attention as a potential solution to this issue. This survey paper presents a systematic and timely review of research efforts on self-supervised recommendation (SSR). Specifically, we propose an exclusive definition of SSR, on top of which we develop a comprehensive taxonomy to divide existing SSR methods into four categories: contrastive, generative, predictive, and hybrid. For each category, we elucidate its concept and formulation, the involved methods, as well as its pros and cons. Furthermore, to facilitate empirical comparison, we release an open-source library SELFRec (\url{https://github.com/Coder-Yu/SELFRec}), which incorporates a wide range of SSR models and benchmark datasets. Through rigorous experiments using this library, we derive and report some significant findings regarding the selection of self-supervised signals for enhancing recommendation. Finally, we shed light on the limitations in the current research and outline the future research directions.
		\end{abstract}
		
		\begin{IEEEkeywords}
			Recommendation, Self-Supervised Learning, Contrastive Learning, Pre-Training, Data Augmentation.
	\end{IEEEkeywords}}

	% make the title area
	\maketitle

	% To allow for easy dual compilation without having to reenter the
	% abstract/keywords data, the \IEEEtitleabstractindextext text will
	% not be used in maketitle, but will appear (i.e., to be "transported")
	% here as \IEEEdisplaynontitleabstractindextext when the compsoc 
	% or transmag modes are not selected <OR> if conference mode is selected 
	% - because all conference papers position the abstract like regular
	% papers do.
	\IEEEdisplaynontitleabstractindextext
	% \IEEEdisplaynontitleabstractindextext has no effect when using
	% compsoc or transmag under a non-conference mode.

	% For peer review papers, you can put extra information on the cover
	% page as needed:
	% \ifCLASSOPTIONpeerreview
	% \begin{center} \bfseries EDICS Category: 3-BBND \end{center}
	% \fi
	%
	% For peerreview papers, this IEEEtran command inserts a page break and
	% creates the second title. It will be ignored for other modes.
	\IEEEpeerreviewmaketitle	
	
	\IEEEraisesectionheading{\section{Introduction}\label{sec:introduction}}
	\IEEEPARstart{R}{e}commender systems have become a vital tool for discovering users' latent interests and preferences, providing delightful user experience, and driving incremental revenue in various online E-commerce platforms \cite{ricci2011introduction}. In recent years, powered by highly expressive deep neural architectures, modern recommender systems \cite{covington2016deep,cheng2016wide,zhou2018deep,chen2020try,yin2015joint} have achieved tremendous success and yielded unparalleled performance. However, deep recommendation models are inherently data-hungry. To take advantage of the deep architecture, an enormous amount of training data is required. Unlike image annotation that can be undertaken by the crowdsourcing, data acquisition in recommender systems is costly as personalized recommendations rely on the data generated by users themselves. Unfortunately, most users typically interact with only a small fraction of the vast number of available items \cite{sarwar2001sparsity}. Consequently, the data sparsity issue bottlenecks deep recommendation models from reaching their full potential \cite{zhang2019deep}.

	Self-supervised learning (SSL) \cite{liu2021self}, emerging as a learning paradigm that can reduce the dependency on manual labels and enables training on massive unlabeled data, recently has received considerable attention. The essential idea of SSL is to extract transferrable knowledge from abundant unlabeled data through well-designed self-supervised tasks (a.k.a. pretext tasks), in which the supervision signals are semi-automatically generated. Due to the ability to overcome the pervasive label insufficiency problem, SSL has been applied to a diverse set of domains including visual representation learning \cite{he2020momentum,grill2020bootstrap,chen2020simple}, language model pre-training \cite{lan2019albert,devlin2018bert}, audio representation learning \cite{oord2018representation}, node/graph classification \cite{qiu2020gcc,velickovic2019deep}, etc, and it has been proved a powerful technique. As the principle of SSL is well-aligned with recommender systems' needs for more annotated data, motivated by the immense success of SSL in other domains, a large and growing body of research is now exploring the application of SSL to recommendation. 

	The early prototypes of self-supervised recommendation (SSR) date back to unsupervised methods like autoencoder-based recommendation models \cite{wu2016collaborative}, which prevent overfitting by relying on different corrupted data to reconstruct the original input. Following this, SSR appeared as network embedding-based recommendation models \cite{gao2018bine,zhang2017collaborative}, which use random-walk proximity as the self-supervision signals to capture similarities between users and items. During the same period, a number of generative adversarial networks \cite{goodfellow2014generative} (GANs)-based recommendation models \cite{wang2019enhancing,yu2020enhance} that augment user-item interactions can be seen as another embodiment of SSR. In 2018, the advent of the groundbreaking pre-trained language model BERT \cite{devlin2018bert} brought SSL, as an independent concept, into the spotlight. The recommendation community soon embraced SSL and started to work on pre-training recommendation models with the Cloze-like tasks on sequential data \cite{sun2019bert4rec,chen2019bert4sessrec,zhou2020s3}. Since 2020, SSL has enjoyed a period of prosperity, with the latest SSL-based methods even demonstrating comparable performance to their supervised counterparts in many computer vision (CV) and natural language processing (NLP) tasks \cite{grill2020bootstrap,goyal2021self}. Particularly, the resurgence of contrastive learning (CL) \cite{jaiswal2021survey} significantly pushes forward the frontier of SSL. Accordingly, a flurry of enthusiasm on SSR has also been witnessed \cite{xia2021aaai,yu2021self,wujc2021self,lee2021bootstrapping,xinxin2020self}. The paradigms of SSR become diverse and the scenarios are no longer limited to sequential recommendation. 
	%For instance, BUIR \cite{lee2021bootstrapping} borrows idea from CV and duplicates the two-branch neural architecture in BYOL \cite{grill2020bootstrap} to bootstrap user and item representations for item recommendation on graphs. MHCN \cite{yu2021self} extends SSR to hypergraphs with hierarchical contrastive learning for social recommendation. 
	\par

	% \begin{figure}[t]
	% 	\centering
	% 	\subfigure[Article numbers (By 2022/02).]{
	% 	\begin{minipage}[t]{.23\textwidth}
	% 	\centering
	% 	\includegraphics[width=1.5in]{articlenum}
	% 	\label{p1.a}
	% 	%\caption{fig1}
	% 	\end{minipage}%
	% 	}%
	% 	\subfigure[Contribution of each venue.]{
	% 	\begin{minipage}[t]{.23\textwidth}
	% 	\centering
	% 	\includegraphics[width=1.5in]{venue}
	% 	\label{p1.b}
	% 	%\caption{fig2}
	% 	\end{minipage}%
	% 	}%
	% 	\centering
	% 	\caption{Statistics of papers on self-supervised recommendation.}
	% \end{figure}

	\par
	While SSL has been extensively surveyed in the fields of CV, NLP \cite{jing2020self,liu2021self} and graph learning \cite{wu2021self,liu2021graph,xie2021self}, there has not been a systematic investigation of research endeavors on SSR despite the growing number of publications. Unlike the aforementioned fields, recommendation involves a plethora of scenarios with varying optimization objectives and multiple types of data, making it difficult to generalize the ready-made SSL methods designed for other domains to recommendation. Meanwhile, recommender systems encounter unique challenges such as highly-skewed data distribution \cite{DBLP:journals/pvldb/YinCLYC12}, widely observed biases \cite{chen2020bias}, and large-vocabulary categorical features \cite{covington2016deep}, which provide soil for new-type SSL and have spurred a series of distinct SSR methods that can enrich the SSL family. Given the increasing prevalence of SSR, there is an urgent need for a timely and systematic survey to summarize the current achievements, discuss the strengths and limitations of existing research efforts on SSR, and promote future research. Therefore, this paper presents an up-to-date and comprehensive retrospective on the frontier of SSR. In summary, our contributions are fourfold:  %conducts a comprehensive and up-to-date overview of the self-supervised recommender systems. The most relevant work to ours is \cite{zeng2021knowledge}, which retrospects the development the pre-trained recommendation models.  
	\begin{itemize}[leftmargin=*]
		\item We present a comprehensive survey of the latest research on SSR, which covers a large number of related papers. To the best of our knowledge, this is the first survey that focuses specifically on SSR.
		\item We provide a unique and precise definition of SSR, along with its connections to related concepts. Moreover, we develop a comprehensive taxonomy that categorizes existing SSR methods into four types: contrastive, generative, predictive, and hybrid. For each category, we discuss its concept and formulation, the involved methods, as well as its strengths and limitations.
		\item We introduce an open-source library, SELFRec, which aims to facilitate the implementation and evaluation of SSR models. The library incorporates multiple benchmark datasets and evaluation metrics, and includes more than 20 state-of-the-art SSR methods. Through rigorous experiments using SELFRec, we derive significant findings regarding designing effective SSR. 
		\item We shed light on the limitations in the existing research, and identify the remaining challenges and future directions to advance SSR.
	\end{itemize}
	\par
	\noindent\textbf{Paper collection}. In this survey, we comprehensively review over 60 high-quality papers that solely focus on SSR and were published after 2018. Prior implementations of SSR, such as autoencoder-based and GAN-based recommendation models, have been extensively covered in previous surveys on deep learning \cite{zhang2019deep,wu2021survey} and adversarial training \cite{gao2020recommender,deldjoo2021survey}. Therefore, we will not revisit them in the ensuing chapters. In conducting our literature search, we utilized DBLP and Google Scholar as the primary search engines with the keywords "self-supervised + recommendation," "contrastive + recommendation," "augmentation + recommendation," and "pre-training + recommendation." We then traversed the citation graph of the identified papers and included relevant studies. Furthermore, we monitored top-tier conferences and journals such as ICDE, CIKM, ICDM, KDD, WWW, SIGIR, WSDM, AAAI, IJCAI, TKDE, TOIS, etc., to ensure that we did not omit important work. In addition to published papers, we also screened preprints on arXiv and identified those with novel and interesting ideas for a more inclusive panorama. %In Fig. \ref{p1.b}, we visualize the percents of collected papers that each venue contributes.
	\par
	\noindent\textbf{Connections to existing surveys}. Although there are some surveys on graph SSL \cite{wu2021self,liu2021graph,xia2022survey} that cover a few papers on recommendation, they just take those works as the supplementary applications of graph SSL. Another relevant survey \cite{zeng2021knowledge} pays attention to the pre-training of recommendation models. However, its focus is transferring knowledge between different domains by exploiting knowledge graphs, and only covers a small number of BERT-like works. Compared with them, our survey purely centers on recommendation-specific SSL and is the first one to provide a systematic review of a large number of up-to-date papers in this line of research. 
	\par
	\noindent\textbf{Targeted audiences}. This survey is expected to provide significant benefits to various stakeholders in the recommendation community. First, researchers and practitioners who are new to the field of SSR will find this survey an efficient way to quickly familiarize themselves with this area. Second, for those who are struggling to navigate the numerous self-supervised approaches, this survey offers a clear pathway. Third, those who are interested in staying up-to-date with the latest developments in SSR will find this survey a valuable resource. Finally, for developers who are currently working on developing SSR, this survey will offer useful guidance and insights.
	\par
	\noindent\textbf{Survey structure}. The remainder of this survey is structured as follows. In section \ref{sec:taxonomy} we begin with the definition and  formulation of SSR, followed by the taxonomy distilled from surveying a large number of research papers. Section \ref{sec:augmentation} introduces the commonly used data augmentation approaches. Sections \ref{sec:contrastive}-\ref{sec:Hybrid} provide a detailed review of the four categories of SSR models, along with their respective advantages and disadvantages. Section \ref{sec:evaluation} introduces the open-source framework SELFRec and Section \ref{sec:experiment} presents the experimental findings derived through using SELFRec. Section \ref{sec:discussion} discusses the limitations in current research and identifies some promising directions for inspiring future research. Finally, section \ref{sec:conclusion} concludes this paper.
	
	\begin{figure}[t]
		\centering
		\captionsetup{justification=centering}
		\includegraphics[width=.48\textwidth]{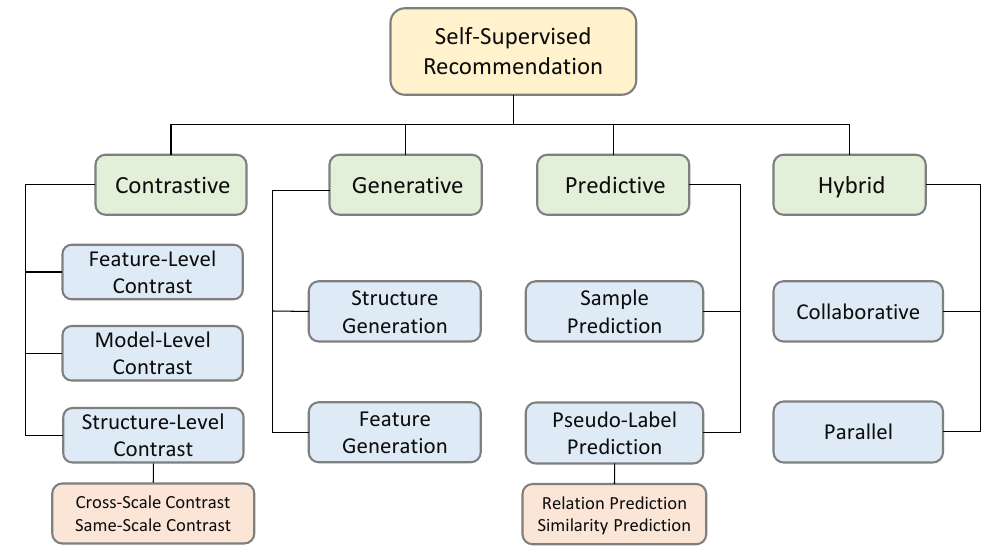}
		\caption{The taxonomy of self-supervised recommendation. }
		\label{fig:taxonomy}				
	\end{figure}
	
	%input external tex files
	\section{Definition and Taxonomy}\label{sec:taxonomy}
In this section, we first define and formalize SSR. Then we lay out a comprehensive taxonomy to categorize existing SSR methods into four paradigms based on the features of their pretext tasks (Fig. \ref{fig:taxonomy}). At last, we introduce three typical training schemes of SSR. 
\subsection{Preliminaries}
The current research of SSR mainly exploits the graph and sequential data, where the original user-item interactions are modeled as bipartite graphs and item sequences in chronological order, respectively. In the scenario of graph-based recommendation, we let $\mathcal{G}=(\mathcal{V},\mathcal{E})$ denote the user-item bipartite graph where $\mathcal{V}$ is the node set (i.e. users $\mathcal{U}$ and items $\mathcal{I}$ ), and $\mathcal{E}$ is the edge set (i.e. interactions). The graph structure is represented with the adjacency matrix $\mathbf{A}$ where $\mathbf{A}_{ui}$=$1$ denotes that node $u$ and node $i$ are connected. In the scenario of sequential recommendation, we let $\mathcal{I}=[i_{1},i_{2},...,i_{n}]$ denote the item set. The behaviors of each user are often modeled as an ordered sequence $S_{u}=[i^{u}_{1},i^{u}_{2},...,i^{u}_{k}],(1\leq k\leq n)$, and $\mathcal{S}=\{S_{1},S_{2},...,S_{m}\}$ refers to the whole dataset. In some cases, the users and items are associated with their attributes. We use $\mathbf{X} = [\mathbf{x}_{1}, \mathbf{x}_{2},..., \mathbf{x}_{m+n}]$ to denote the attribute matrix where $\mathbf{x}_{i}\in{\mathbb{R}^{t}}$ is the multi-hot vector representing object $i$'s attributes. The general purpose of recommendation models is to learn quality representations $\mathbf{H}\in\mathbb{R}^{(m+n)\times d}=[\mathbf{U},\mathbf{V}]$ for the users and items to generate satisfactory recommendation results, where $d$ is the dimension of representations. To facilitate the reading, in this paper, matrices are denoted by capital letters, vectors appear in bold lowercase letters, sets are represented with italic capital letters. 

\subsection{Definition and Formulation}\label{subsec:definition}
SSL provides a new way to conquer the data sparsity issue in recommendation. However, there is currently no formal definition of SSR. In order to establish a solid foundation for subsequent research in this area, we propose a clear and accurate definition of SSR by examining the collected literature, with its three key features summarized as follows:
\begin{enumerate}[label=(\roman*)] \label{definition}
    \item Semi-automatically exploiting the raw data itself to obtain more supervision signals.
    \item Incorporating a self-supervised task(s) to (pre-)train the recommendation model using augmented data.
    \item The self-supervised task is designed to enhance recommendation performance, rather than being an end goal.
\end{enumerate}
Of these features, (\rmnum{1}) is the fundamental premise and specifies the \textbf{scope} of SSR. By leveraging the raw data itself rather than requesting more data, SSR aims to extract additional supervision signals to complement the sparse explicit feedback. (\rmnum{2}) describes the \textbf{setup} of SSR, which is a key differentiator from traditional recommendation models. Here augmented data refers to new training examples generated by applying various transformations to original data (e.g. a perturbed graph with its edges dropped at a certain rate) and self-supervised tasks refer to the process wherein the augmented data is generated and exploited (e.g., structure generation with corrupted features from neighboring nodes). The incorporation of self-supervised tasks and augmented data is a prerequisite for SSR. (\rmnum{3}) highlights the \textbf{primary \& auxiliary relation} between the recommendation task and the self-supervised task.

%In the surveyed papers, the raw data, in most cases, only refers to the user-item interactions and sometimes includes the associated attributes because most open datasets for academic use only contain these information. 
The proposed definition allows us to differentiate SSR from related recommendation approaches. For example, pre-training-based recommendation~\cite{zeng2021knowledge} is often conflated with SSR due to the prevalence of pre-training as a standard SSL technique in other domains. However, some pre-training-based recommendation methods~\cite{meng2021graph,moreira2021transformers} are purely supervised, lacking data augmentation and requiring additional human-annotated side information for pre-training. As a result, the two paradigms are only partially overlapped, and should not be treated as synonymous. Analogously, contrastive learning (CL)~\cite{jaiswal2021survey} based recommendation is often considered equivalent to self-supervised recommendation. However, CL can be applied to both supervised and unsupervised settings, and those CL-based recommendation methods which do not augment the raw data~\cite{sankar2020groupim,wei2021contrastive} and just optimize a marginal loss~\cite{qin2021world,zhou2021contrastive}, should not be roughly classified into SSR either. 
\par
Given the diverse data types and optimization objectives in recommender systems, a model-agnostic framework is necessary to formulate SSR. While the specific structures and number of encoders and projection heads may vary across models, most existing approaches can be sketched into an \textbf{Encoder}\ + \textbf{Projection-Head} architecture. To accommodate different data modalities, such as graphs, sequences, and categorical features, a range of neural networks, such as Graph Neural Networks (GNNs) \cite{wu2020comprehensive}, Transformers \cite{vaswani2017attention}, and Multi-Layer Perceptrons (MLPs), can be employed as the encoder $f_{\theta}$, while the projection head $g_{\phi}$ (also referred to as the decoder in generative models) is typically a lightweight structure, such as a linear transformation, a shallow MLP, or a non-parametric mapping. The encoder $f_{\theta}$ aims to learn distributed representations $\mathbf{H}$ for users and items, while the projection head $g_{\phi}$ refines $\mathbf{H}$ for either the recommendation task or a specific self-supervised task. Based on this architecture, SSR can be formulated as follows:
\begin{equation}
f_{\theta^{*}},g_{\phi^{*}}, \mathbf{H}^{*} = \mathop{\arg\min}\limits_{f_{\theta}, g_{\phi}}\mathcal{L}\left( g_{\phi}(f_{\theta}(\mathcal{D},\tilde{\mathcal{D}})) \right),
\label{eq: unified form}
\end{equation}
where $\mathcal{D}$ denotes the original data, $\tilde{\mathcal{D}}$ refers to the augmented data that satisfies $\tilde{\mathcal{D}} \sim \mathcal{T}(\mathcal{D})$, $\mathcal{T} (\cdot)$ denotes the augmentation module, and $\mathcal{L}$ is the merged loss function that can be divided into the loss of the recommendation task $\mathcal{L}_{rec}$ and the loss of the pretext task $\mathcal{L}_{ssl}$. By minimizing Eq. (\ref{eq: unified form}), the optimal encoder(s) $f_{\theta^{*}}$, projection head(s) $g_{\phi^{*}}$, and representations $\mathbf{H}^{*}$ can be learned for generating quality recommendation results. 

\begin{figure}[t]
    \centering
    \subfigure[Split by self-supervised tasks.]{
    \begin{minipage}[t]{.25\textwidth}
    \centering
    \includegraphics[width=1.6in]{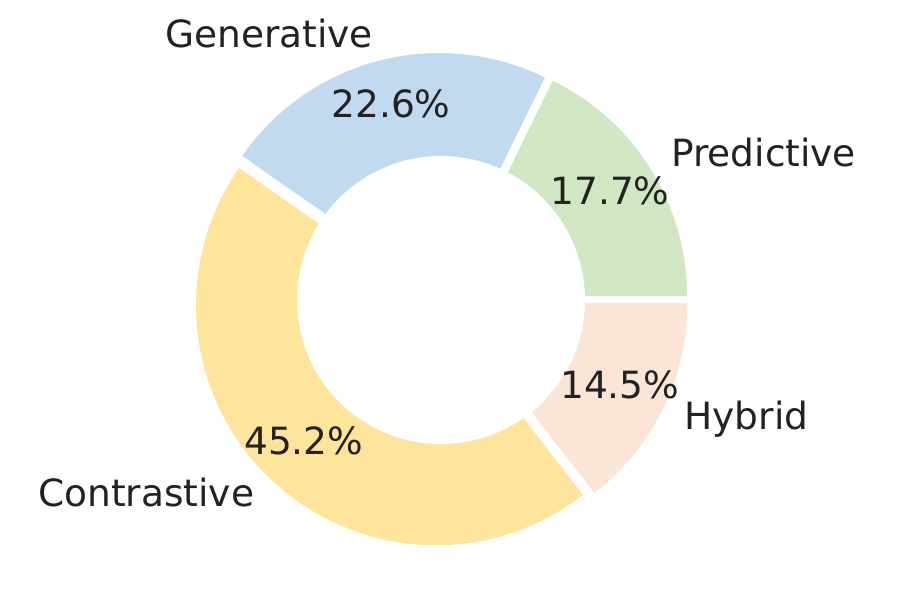}
    \label{fig:category}
    %\caption{fig1}
    \end{minipage}%
    }%
    \subfigure[Split by training schemes.]{
    \begin{minipage}[t]{.25\textwidth}
    \centering
    \includegraphics[width=1.6in]{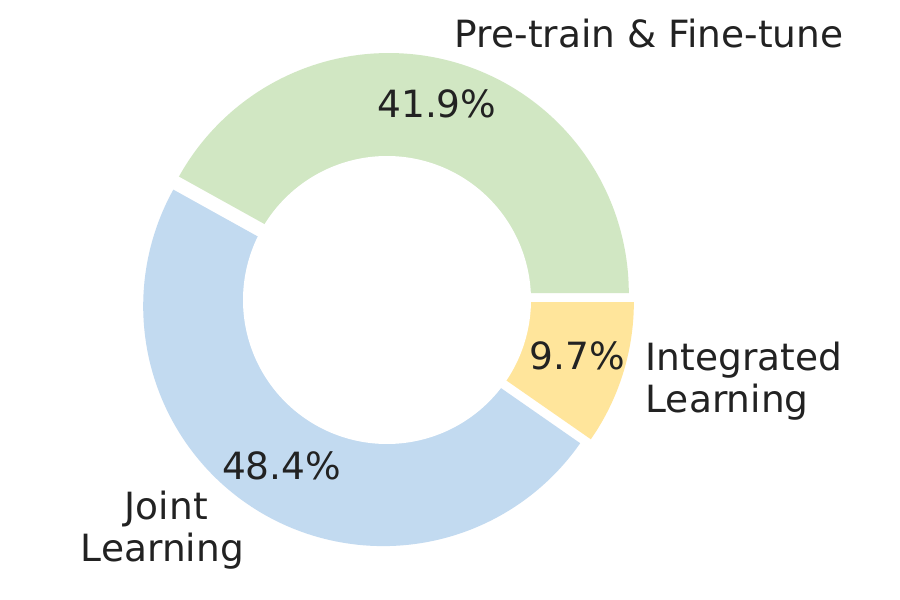}
    \label{fig:training}
    %\caption{fig2}
    \end{minipage}%
    }%
    \centering
    \caption{Distributions of self-supervised recommendation.}
\end{figure}

\subsection{Taxonomy} \label{subsec: categorization}
SSR is distinguished from other recommendation paradigms by emphasizing the role of self-supervised tasks in its methodology. We divide the existing SSR models into four categories: contrastive, predictive, generative, and hybrid according to the nature of their self-supervised tasks.
\subsubsection{Contrastive Methods} 
Driven by CL \cite{jaiswal2021survey}, contrastive methods have become the dominant branch in SSR (shown in Fig. \ref{fig:category}). The fundamental idea behind contrastive methods is to treat every instance (e.g., user/item/sequence) as a class, and then pull variants of the same instance closer in the embedding space, and push variants of different instances apart, where the variants are created by imposing different transformations on the original data (Fig. \ref{fig: contrastive}). Generally, two variants of the same instance are considered a positive pair, and variants of different instances are considered negative samples of each other. A variant is supposed to introduce the non-essential variations rather than significantly modify the original instance. By maximizing the consistency between positive pairs while minimizing the agreement between negative pairs, the method can obtain discriminative representations for recommendation. We formulate the contrastive task as:
\begin{equation}
    f_{\theta}^{*}=\mathop{\arg\min}\limits_{f_{\theta}, g_{\phi_{s}}}\mathcal{L}_{ssl}\Big( g_{\phi_{s}}(f_{\theta}(\tilde{\mathcal{D}}_{1}), f_{\theta}(\tilde{\mathcal{D}}_{2}))\Big),
\label{eq: contrastive}
\end{equation}
where $\tilde{\mathcal{D}}_{(1)}\sim \mathcal{T}_{1}(\mathcal{D})$ and $\tilde{\mathcal{D}}_{(2)}\sim \mathcal{T}_{2}(\mathcal{D})$ are two differently augmented variants of $\mathcal{D}$, $\mathcal{T}_{1}(\cdot)$ and $\mathcal{T}_{2}(\cdot)$ are augmentation operators. The loss function $\mathcal{L}_{ssl}$ estimates the mutual information (MI) between examples through the representations learned by the shared encoder $f_{\theta}$.

\subsubsection{Generative Methods}
Generative approaches draw inspiration from masked language models (MLM) such as BERT \cite{devlin2018bert}. These models employ a self-supervised task wherein the original user/item profile is reconstructed from its corrupted versions (Fig. \ref{fig: generative}). The model is trained to predict a proportion of the available data from the rest, with structure and feature reconstruction being the most common tasks. The self-supervised task is typically formulated as:
\begin{equation}
    f_{\theta}^{*}=\mathop{\arg\min}\limits_{f_{\theta}, g_{\phi_{s}}}\mathcal{L}_{ssl}\Big( g_{\phi_{s}}\big( f_{\theta}(\tilde{\mathcal{D}})\big) , \mathcal{D} \Big),
\label{eq: generative}
\end{equation}
where $\tilde{\mathcal{D}}\sim \mathcal{T}(\mathcal{D})$ denotes the corrupted version of the original input. For most of the generative SSR methods, the objective function $\mathcal{L}_{ssl}$ is often instantiated as either the Cross Entropy (CE) loss or the Mean Squared Error (MSE) which estimates the probability distribution/values of the masked items/numerical features.

\subsubsection{Predictive Methods}
\begin{figure}[t]
    \centering  
    \subfigure[Contrastive Methods]{
    \begin{minipage}[t]{0.5\textwidth}
    \centering
    \label{fig: contrastive}
    \includegraphics[width=3.4in]{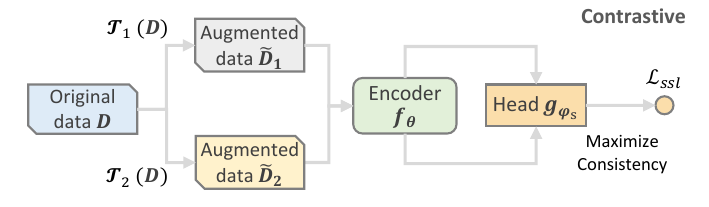}
    %\caption{fig1}
    \end{minipage}%
    }%
    
    \subfigure[Generative Methods]{
    \begin{minipage}[t]{0.5\textwidth}
    \centering
    \label{fig: generative}
    \includegraphics[width=3.4in]{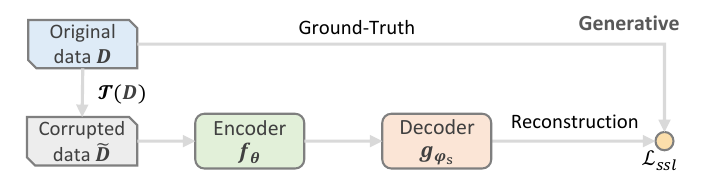}
    %\caption{fig2}
    \end{minipage}%
    }%
    
    \subfigure[Predictive Methods]{
    \begin{minipage}[t]{0.5\textwidth}
    \centering
    \label{fig: predictive}    
    \includegraphics[width=3.4in]{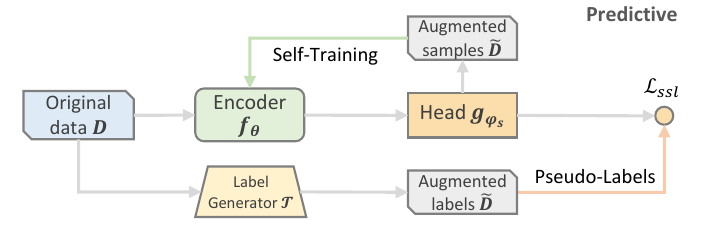}
    %\caption{fig2}
    \end{minipage}%
    }

    \subfigure[Hybrid Methods]{
    \begin{minipage}[t]{0.5\textwidth}
    \centering
    \label{fig: hybrid}    
    \includegraphics[width=3.4in]{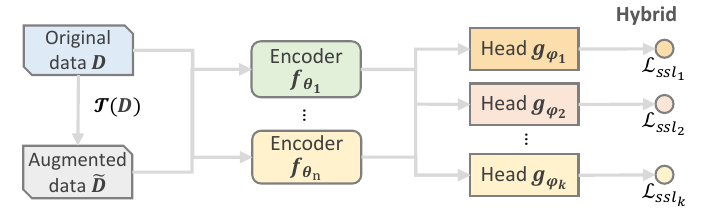}
    %\caption{fig2}
    \end{minipage}%
    }
    \caption{Four common paradigms of self-supervised recommendation.}
    \label{fig: ssr categorization}
    \end{figure}

Predictive methods in SSR may appear similar to generative methods as both involve prediction, but the underlying objectives are distinct. Generative methods focus on predicting missing parts of the original data, which can be viewed as a form of self-prediction. In contrast, predictive methods generate new samples or labels from the original data to guide the pretext task. We classify existing predictive SSR methods into two categories: sample-based and pseudo-label-based (Fig. \ref{fig: predictive}). Sample-based methods aim to predict informative samples based on the current encoder parameters. These predicted samples are then fed back into the encoder to generate new samples with higher confidence. This approach connects self-training, which is a flavor of semi-supervised learning, and SSL. Pseudo-label-based methods, on the other hand, generate labels through a generator, which can be another encoder or rule-based selectors. These generated labels are then used as ground truth to guide the encoder $f_{\theta}$. The pseudo-label-based approach can be formulated as follows:
\begin{equation}
    f_{\theta}^{*}=\mathop{\arg\min}\limits_{f_{\theta}, g_{\phi_{s}}}\mathcal{L}_{ssl}\Big( g_{\phi_{s}}\big( f_{\theta}(\mathcal{D})\big) , \tilde{\mathcal{D}} \Big),
    \label{eq: predictive}
\end{equation}
where $\tilde{\mathcal{D}}\sim \mathcal{T}(\mathcal{D})$ denotes the generated labels, and $\mathcal{L}_{ssl}$ often appears in forms of the CE/softmax or MSE. The former aligns the predicted probability with the labels and the latter measures the difference between the output of $g_{\phi_{s}}$ and the labels, which correspond to the classification problem and the regression problem, respectively.

\subsubsection{Hybrid Methods}
Each type of methods described above has its own distinct advantages and can leverage different self-supervision signals. A viable strategy to obtain comprehensive self-supervision is to combine various self-supervised tasks and integrate them into a single recommendation model. Hybrid methods typically require multiple encoders and projection heads (\ref{fig: hybrid}), and different self-supervised tasks may operate in parallel or collaborate to enhance self-supervision signals. The combination of various pretext tasks is typically formulated as a weighted sum of different self-supervised losses presented in the aforementioned categories.

\subsection{Typical Training Schemes}\label{subsec: training schemes}
Although SSR has a unified formulation (Eq. (\ref{eq: unified form})), the recommendation task is coupled with the pretext task in distinct ways in various scenarios. In this section, we present three typical training schemes of SSR: Joint Learning (JL), Pre-training and Fine-tuning (PF), and Integrated Learning (IL).

\subsubsection{Joint Learning (JL)}
As depicted in Figure \ref{fig:training}, nearly half of the collected SSR methods prefer the joint learning training scheme wherein the pretext task and the recommendation task are jointly optimized with a shared encoder (Figure \ref{fig: training scheme.b}). The balance between the two objectives $\mathcal{L}_{ssl}$ and $\mathcal{L}_{rec}$ is achieved by tuning the hyper-parameter $\alpha$, which controls the level of self-supervision. While the JL scheme can be regarded as a form of multi-task learning, the output of the pretext task is typically not prioritized and is instead considered an auxiliary task that helps to regularize the recommendation task. The formulation of the JL scheme is as follows:
\begin{equation}
        \mathbf{\Theta}^{*} = \mathop{\arg\min}\limits_{f_{\theta}, g_{\phi}}\mathcal{L}_{rec}\big(g_{\phi_{r}}(f_{\theta}(\mathcal{D}))\big)
        +\alpha\mathcal{L}_{ssl}\big(g_{\phi_{s}}(f_{\theta}(\tilde{\mathcal{D}}))\big)
\label{eq: JL}
\end{equation}
For the sake of brevity, we use $\mathbf{\Theta}^{*}$ to denote all the parameters. This scheme is mostly used in contrastive methods.

\subsubsection{Pre-training and Fine-tuning (PF)}
The PF scheme is the second most commonly used training scheme, comprising two stages: pre-training and fine-tuning (Figure \ref{fig: training scheme.a}). In the pre-training stage, the encoder $f_{\theta}$ is pre-trained with the self-supervised task on augmented data to achieve a favorable initialization of its parameters. Subsequently, $f_{\theta_{init}}$ is fine-tuned on the original data, followed by a projection head $g_{\phi_{r}}$ for the recommendation task. Prior studies on graphs \cite{wu2021self,xie2021self} have introduced another training scheme known as unsupervised representation learning. This scheme first pre-trains the encoder then freezes it, only learning a small number of additional parameters for downstream tasks. We consider this approach to be a special case of the PF scheme and it only appears in the training of general-purpose recommendation models \cite{liu2023pre,hou2022towards}. The formulation of the PF scheme is defined as follows:
\begin{equation}
\begin{aligned}
f_{\theta_{init}} &= \mathop{\arg\min}\limits_{f_{\theta}, g_{\phi_{s}}}\mathcal{L}_{ssl}\big( g_{\phi_{s}}(f_{\theta}(\tilde{\mathcal{D}}),\mathcal{D})\big) \\
\mathbf{\Theta}^{*} &= \mathop{\arg\min}\limits_{f_{\theta_{init}}, g_{\phi_{r}}}\mathcal{L}_{rec}\big(g_{\phi_{r}}(f_{\theta}(\mathcal{D}))\big)
\end{aligned}
\label{eq: PT&FT}
\end{equation}
This scheme is commonly utilized to train BERT-like generative SSR models. Additionally, some contrastive methods also leverage this training scheme, where the contrastive pretext task is employed for pre-training.

\subsubsection{Integrated Learning (IL)}
\begin{figure}[t]
    \centering
    \subfigure[Joint Learning (JL)]{
        \begin{minipage}[t]{0.5\textwidth}
        \centering
        \label{fig: training scheme.b}
        \includegraphics[width=3.4in]{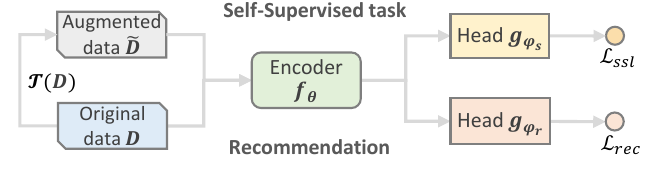}
        %\caption{fig2}
        \end{minipage}%
    }%

    \subfigure[Pre-train and Fine-tune (PF)]{
    \begin{minipage}[t]{0.5\textwidth}
    \centering
    \label{fig: training scheme.a}
    \includegraphics[width=3.4in]{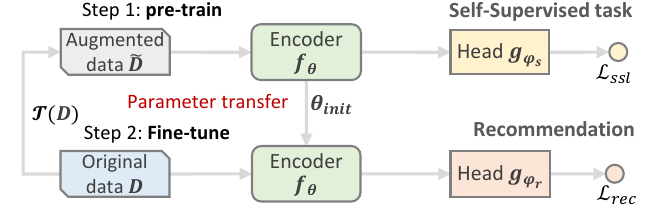}
    %\caption{fig1}
    \end{minipage}%
    }%
        
    \subfigure[Integrated Learning (IL)]{
    \begin{minipage}[t]{0.5\textwidth}
    \centering
    \label{fig: training scheme.c}    
    \includegraphics[width=3.4in]{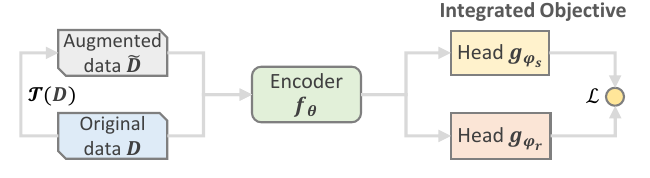}
    %\caption{fig2}
    \end{minipage}%
    }
    \caption{Three typical training schemes of self-supervised pretext tasks.}
    \label{fig: training scheme}
    \end{figure}

In comparison to the JL and PF schemes, the IL scheme has received less attention and is not widely adopted (Figure \ref{fig: training scheme.c}). In this scheme, the pretext task and the recommendation task are well-aligned and integrated into a unified objective. The loss function $\mathcal{L}$ typically quantifies the dissimilarity or mutual information between the two outputs. The IL scheme can be formulated as follows:
\begin{equation}
\begin{aligned}
    \mathbf{\Theta}^{*} &= \mathop{\arg\min}\limits_{f_{\theta}, g_{\phi}}\mathcal{L}\left(g_{\phi_{r}}(f_{\theta}(\mathcal{D})), g_{\phi_{s}}(f_{\theta}(\tilde{\mathcal{D}}))\right)
\end{aligned}
\label{eq: URL}
\end{equation}
This scheme is mainly used by the pseudo-labels-based predictive methods and a few contrastive methods.

	\section{Data Augmentation} \label{sec:augmentation}
Prior studies \cite{ding2022data,li2021data,tian2020makes} have highlighted the crucial role played by data augmentation in facilitating the learning of high-quality, generalizable representations. Before delving into SSR methods, we present an overview of commonly used data augmentation techniques in SSR and classify them into three categories: sequence-based, graph-based, and feature-based. Most of these augmentation methods are task-independent and model-agnostic and have been employed across various paradigms of SSR. For task- and model-dependent approaches, we will introduce them along with specific SSR methods in Sections \ref{sec:contrastive}-\ref{sec:Hybrid}.

\subsection{Sequence-Based Augmentation} \label{data augmentation}
\begin{figure}[t]
    \centering  
    \subfigure[Sequence-based data augmentation]{
    \begin{minipage}[t]{0.5\textwidth}
    \centering
    \label{fig: sequence-augmentation}
    \includegraphics[width=3.4in]{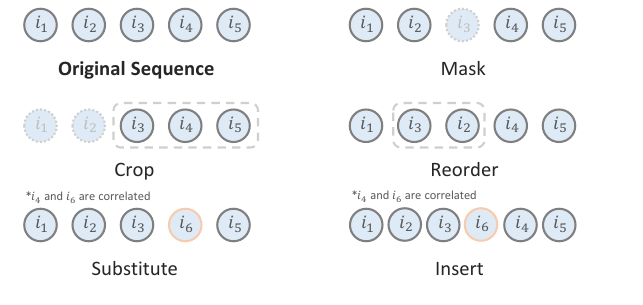}
    %\caption{fig1}
    \end{minipage}%
    }%
    
    \subfigure[Graph-based data augmentation]{
    \begin{minipage}[t]{0.5\textwidth}
    \centering
    \label{fig: graph-augmentation}
    \includegraphics[width=3.4in]{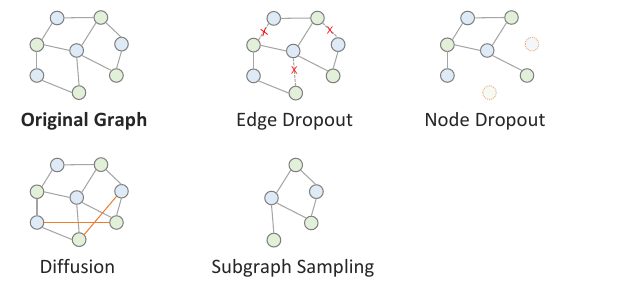}
    %\caption{fig1}
    \end{minipage}%
    }%
    \caption{Data augmentations on sequences and graphs.}
    \label{fig: data-augmentation}
    \end{figure}

Given a sequence of items $S=[i_{1},i_{2},...,i_{k}]$, the common sequence-based augmentations (Fig. \ref{fig: sequence-augmentation}) include:

 \textbf{Item Masking}. Analogous to the word masking in BERT \cite{devlin2018bert}, the item masking strategy used in \cite{cheng2021learning,zhou2020s3,xie2020contrastive} randomly masks a proportion $\gamma$ of items and replaces them with special tokens [$\mathrm{mask}$]. The idea behind is that a user's intention is relatively stable during a period of time. Therefore, though part of items are masked, the primary intent information is still retained in the rest. This augmentation approach can be formulated as:
\begin{equation}
    \begin{aligned}
    \tilde{S} =\mathcal{T}_{\text {masking}}(S)=[\tilde{i}_{1}, \tilde{i}_{2},..., \tilde{i}_{k}],
    \tilde{i}_{t} &= \begin{cases}i_{t}, & t\notin\mathcal{M} \\
    {[\text {mask}],} & t\in\mathcal{M}\end{cases}
    \end{aligned}
\end{equation}

\textbf{Item Cropping}. Inspired by the image cropping in CV, some works \cite{xie2020contrastive,zhou2020s3,li2021hyperbolic,cheng2021learning} propose the item cropping method. Given a user's historic sequence $S$, a continuous sub-sequence with length $L_{\mathrm{c}}=\lfloor\eta *|S|\rfloor$ is randomly chosen, where $\eta\in(0,1)$ is a coefficient that adjusts the length. This method can be formulated as:
\begin{equation}
    \tilde{S}=\mathcal{T}_{\text {cropping}}(S)=[\tilde{i}_{c}, \tilde{i}_{c+1}, \ldots, \tilde{i}_{c+L_{c}-1}]\\
\end{equation}
This method provides a local view of the user's historic sequence. Through the self-supervised task on the augmented data, it is expected that the model can learn generalized representations without the comprehensive user profile. 

\textbf{Item Reordering}. Item transitions in sequences are often assumed strictly context-dependent. However, this assumption is problematic because in the real world many unobserved factors can impact the item order and different item orders may actually correspond to the same user intent. %It is necessary to train the model to be robust to unobserved sequences by encouraging the model to rely less on the item orders. 
Some works \cite{xie2020contrastive,cheng2021learning} propose to shuffle a continuous sub-sequence $[i_{r},i_{r+1},...,i_{r+L_{r}-1}]$ to $[\tilde{i}_{r},\tilde{i}_{r+1},...,\tilde{i}_{r+L_{r}-1}]$ to create sequence augmentations, which is formulated as:
\begin{equation}
    \tilde{S}=\mathcal{T}_{\text {reordering}}(S)=[i_{1},...,\tilde{i}_{r},\tilde{i}_{r+1},...,\tilde{i}_{r+L_{r}-1},...,i_{k}]
\end{equation}

\textbf{Item Substitution}. Random item cropping and masking may exaggerate the data sparsity issue in short sequences. \cite{liu2021contrastive} proposes to substitute items in short sequences with highly correlated items, which injects less corruption to the original sequential information. Given $\mathcal{Z}$ denoting the indices of the randomly selected items which are going to be substituted, the item substitution is formulated as:
\begin{equation}
    \begin{aligned}
        \tilde{S} &=\mathcal{T}_{\text {substitution}}(S)=[\tilde{i}_{1}, \tilde{i}_{2},..., \tilde{i}_{k}],\\
    \tilde{i}_{t} &= \begin{cases}i_{t}, & t\notin\mathcal{Z} \\
    {\text {item correlated to } i_{t},} & t\in\mathcal{Z}\end{cases}
    \end{aligned}
\end{equation}
where the correlated item is obtained by calculating the correlation score which is based on the item co-occurrence or the similarity of the corresponding representations.

\textbf{Item Insertion}. To deal with short sequences, \cite{liu2021contrastive} also proposes to insert correlated items to complement the sequence. $l$ items $[id_{1}, id_{2},..., id_{l}]$ are firstly selected from the given sequence at random, and then their highly correlated items are inserted around them. After the insertion, the augmented sequence with length $l+k$ is:
\begin{equation}
    \tilde{S}=\mathcal{T}_{\text {insert}}(S)=[i_{1},...,\tilde{i}_{id_{1}},i_{id_{1}},...,\tilde{i}_{id_{l}},i_{id_{l}}...,i_{k}]
\end{equation}
 %Some works \cite{liu2021augmenting,jiang2021sequential} propose to predict the prior items which are inserted at the beginning of the original sequences in a right-to-left pre-training fashion. 

\subsection{Graph-Based Augmentation}
Given the user-item graph $\mathcal{G}=(\mathcal{V},\mathcal{E})$ with the adjacency matrix $\mathbf{A}$ (or other graphs like the user-user graph), the following augmentation approaches (Fig. \ref{fig: graph-augmentation}) can be applied.

\textbf{Edge/Node Dropout}. With a probability $\rho$, each edge may be removed from the graph. The idea behind is that only partial connections contribute to the node representations, and discarding redundant connections can endow the representations more robustness, which is analogous to the cropping methods. This method is formulated as:
\begin{equation}
    \tilde{\mathcal{G}},\tilde{\mathbf{A}}=\mathcal{T}_{\text {E-dropout}}(\mathcal{G})=(\mathcal{V},\mathbf{m}\odot\mathcal{E}),
\end{equation}
where $\mathbf{m}\in(0,1)^{|\mathcal{E}|}$ is the masking vector on the edge set generated by a Bernoulli distribution. This method is widely used in many contrastive methods \cite{lee2021bootstrapping,zhang2021double,wujc2021self,wang2021curriculum,yu2021kdd,zhou2021selfcf}. 
Similarly, each node can also be dropped from the graph, together with its associated edges \cite{hao2021multi,wujc2021self,wang2021curriculum}. This augmentation method is expected to identify the influential nodes from differently augmented views, which is formulated as:
\begin{equation}
    \tilde{\mathcal{G}},\tilde{\mathbf{A}}=\mathcal{T}_{\text {N-dropout}}(\mathcal{G})=(\mathcal{V}\odot\mathbf{m},\mathcal{E}\odot\mathbf{m}^{\prime}),
\end{equation}
where $\mathbf{m}\in(0,1)^{|\mathcal{V}|}$ is the masking vector on the node set and $\mathbf{m}^{\prime}$ is the vector masking the associated edges.

\textbf{Graph Diffusion}. Opposite to the dropout-based methods, the diffusion-based augmentation adds edges into graphs to create views. \cite{yang2021enhanced} considers that the missing user behaviors include unknown positive preferences which can be represented with weighted user-item edges. Therefore, they discover the possible edges by calculating the similarities of the user and item representations and retain the edges with top-$K$ similarities. This method is formulated as:
\begin{equation}
    \tilde{\mathcal{G}},\tilde{\mathbf{A}}=\mathcal{T}_{\text {diffusion}}(\mathcal{G})=(\mathcal{V},\mathcal{E}+\tilde{\mathcal{E}})
\end{equation}
When the edges are randomly added, this method can also be used to generate negative samples.

\textbf{Subgraph Sampling}. This method samples a portion of nodes and edges to form subgraphs. Many approaches can be used to induce subgraphs like meta-path guided random walks \cite{yang2021hyper,wang2021curriculum}, and the ego-network sampling \cite{yu2021self,cao2021bipartite,xie2021contrastive,wang2021pre}. The underlying idea of subgraph sampling is analogous to the idea of edge dropout whereas the subgraph sampling usually operates on local structures. Give the sampled node set $\mathbf{\mathcal{Z}}$, this method can be formulated as: 
\begin{equation}
    \tilde{\mathcal{G}},\tilde{\mathbf{A}}=\mathcal{T}_{\text {sampling}}(\mathcal{G})=(\mathcal{Z}\in\mathcal{V},\mathbf{A}[\mathcal{Z},\mathcal{Z}])
\end{equation}

\subsection{Feature-Based Augmentation}
Feature-based augmentations operate within the attribute or embedding space. Within this context, categorical attributes and learned continuous embeddings are referred to as "features" and are denoted as $\mathbf{X}$ for simplicity.

\textbf{Feature Dropout} \cite{zhou2020s3,zhou2021selfcf,yao2021self,Qiuwsdm22,qiu2021memory,liu2021selficlr} is similar to the edge dropout, which randomly drops a small portion of features and is formulated as:
\begin{equation}
    \tilde{\mathbf{X}}=\mathcal{T}_{\text {F-dropout}}(\mathbf{X})=\mathbf{X}\odot\mathbf{M},
\end{equation}
where $\mathbf{M}$ is the masking matrix that $\mathbf{M}_{i,j}=0$ if the $j$-th element of vector $i$ is masked/dropped, otherwise $\mathbf{M}_{i,j}=1$. The matrix $\mathbf{M}$ is generated by Bernoulli distribution.

\textbf{Feature Shuffling} \cite{yu2021self,xia2021aaai,xie2021adversarial} switches rows and columns in the feature matrix $\mathbf{X}$. By randomly changing the contextual information, $\mathbf{X}$ is corrupted to yield augmentations. This method can be formulated as:
\begin{equation}
    \tilde{\mathbf{X}}=\mathcal{T}_{\text {shuffling}}(\mathbf{X})=\mathbf{P}_{r}\mathbf{X}\mathbf{P}_{c},
\end{equation}
where $\mathbf{P}_{r}$ and $\mathbf{P}_{c}$ are permutation matrices that have exactly one entry of 1 in each row/column and 0s elsewhere.

\textbf{Feature Clustering} proposed by \cite{chen2022intent,lin2022improving,li2020prototypical} combines CL with clustering, assuming the existence of prototypes in the feature/representation space, and each user/item representation should be semantically similar to assigned prototypes where the prototypes are learned through unsupervised learning like EM algorithm. It is formulated as:
\begin{equation}
    \tilde{\mathbf{C}}=\mathcal{T}_{\text {clustering}}(\mathbf{X})=\text{EM}(\mathbf{X},\mathcal{C}),
\end{equation}
where $\mathcal{C}$ is the presupposed clusters (prototypes) and $\tilde{\mathbf{C}}$ is the augmented prototype representations.

\textbf{Feature Mixing} \cite{huang2021mixgcf,zhou2021selfcf} mixes the original user/item features with features from other users/items or previous versions to synthesize informative negative/positive examples \cite{KalantidisSPWL20}. It usually interpolates two samples in the following way:
\begin{equation}
    \tilde{\mathbf{x}}_{i}=\mathcal{T}_{\text {mixing}}(\mathbf{x}_{i})=\alpha\mathbf{x}_{i}+(1-\alpha)\mathbf{x}_{j}^{'},
\end{equation}
where $\alpha\in [0,1]$ is the mixing coefficient that controls the proportion of information from $\mathbf{x}_{i}$.

\textbf{Feature Perturbation} \cite{yu2021graph,yu2022xsimgcl} adds random noises to original user/item representations. As the magnitude of added noises is very small, the augmented representation retains most information of the original representation while introducing some differences. This method is formulated as:
\begin{equation}
    \tilde{\mathbf{x}}_{i}=\mathcal{T}_{\text {perturbation}}(\mathbf{x}_{i})=\mathbf{x}_{i}+\lambda\Delta_{i},
\end{equation}
where $\lambda$ controls the influence of random noises $\Delta_{i}$.

	\section{Contrastive Methods}\label{sec:contrastive}
The various data augmentation approaches and data types spawn diverse forms of contrastive pretext tasks. Based on the origins of self-supervision signals, these tasks can be categorized into three groups: \textbf{structure-level contrast}, \textbf{feature-level contrast}, and \textbf{model-level contrast}. A summary of the surveyed contrastive methods can be found in Table \ref{Table: contrastive methods}.

\subsection{Structure-Level Contrast}
User behavior data is often represented as graphs or sequences, where slight perturbations to the graph/sequence structures may result in similar semantics. By contrasting different structures, shared invariances to structural perturbations can be obtained as self-supervision signals. We follow the taxonomy proposed by \cite{wu2021self,liu2021graph} and divide structure-level contrast into two categories: \textbf{same-scale contrast} and \textbf{cross-scale contrast}. Same-scale contrast involves views from two objects at the same scale, and is further divided into two levels: \textbf{local-local} and \textbf{global-global}. Cross-scale contrast involves views from two objects at different scales and is further divided into \textbf{local-global} and \textbf{local-context}. In graph structures, \textit{local} refers to nodes, and \textit{global} refers to graphs, while in sequence structures, \textit{local} refers to items and \textit{global} refers to sequences.

\subsubsection{Local-Local Contrast}
This type of contrast comes with graph-based SSR models to maximize the mutual information between user/item node representations, which is formulated as:
\begin{equation}
	f_{\theta}^{*}=\mathop{\arg\min}\limits_{f_{\theta}, g_{\phi_{s}}}\mathcal{L}_{\mathcal{MI}}\big(g_{\phi_{s}}(\tilde{\mathbf{h}}_{i},\tilde{\mathbf{h}}_{j})\big),     
\end{equation}
where $\tilde{\mathbf{h}}_{i}$ and $\tilde{\mathbf{h}}_{j}$ are node representations learned from two augmented views via the shared encoder $f_{\theta}$, and $\mathcal{L}_{\mathcal{MI}}$ is the contrastive loss which will be introduced in Section \ref{sec:cl loss}.  

For local level contrast, dropout-based augmentations are the most preferred methods to create perturbed local views. \textbf{SGL} \cite{wujc2021self}, as a representative model, applies node dropout, edge dropout, and random walk augmentations to the user-item bipartite graph. It generates two augmented graphs with the same type of augmentation operator, and learns node embeddings using a shared graph LightGCN encoder $f_{\theta}$ \cite{he2020lightgcn}. The node-level contrast is optimized using the InfoNCE loss \cite{oord2018representation} with in-batch negative sampling, and is jointly optimized with the Bayesian personalized ranking (BPR) loss \cite{rendle2009bpr} for recommendation. Similarly, \textbf{DCL} \cite{liu2021contrastiveL} employs stochastic edge dropout to perturb the $L$-hop ego-network of a node, resulting in two augmented neighborhood subgraphs. It then maximizes agreement between node representations learned on the two subgraphs. \textbf{HHGR} \cite{zhang2021double} proposes a double-scale node dropout method for group recommendation \cite{yin2019social}. The method includes coarse-grained and fine-grained dropout schemes that remove a portion of user nodes from all groups and only drop randomly selected member nodes from a specific group, respectively. It then maximizes the mutual information between the user node representations learned from these two views with different dropout granularities. Besides, \textbf{KGCL} \cite{yang2022knowledge} applies dropout to knowledge graphs and proposes a knowledge-aware contrastive method which contrasts node representations learned from augmentations of the user-item graph and know graphs. 

Subgraph sampling is another popular method for creating local-level graph contrast. \textbf{CCDR} \cite{xie2021contrastive} applies contrastive learning to cross-domain recommendation, using two types of contrastive tasks: intra-CL and inter-CL. The intra-CL task is similar to the contrastive task in DCL \cite{liu2021contrastiveL}, conducted in the target domain using a graph attention network \cite{VelickovicCCRLB18} as the encoder. The inter-CL task aims to maximize the mutual information between representations of the same object learned in the source and target domains. A concurrent work \textbf{PCRec} \cite{wang2021pre} also connects cross-domain recommendation with CL, sampling $r$-hop ego-networks using random walks to augment data. It pre-trains a GIN \cite{XuHLJ19} encoder in the source domain by contrasting sampled subgraphs, and then fine-tunes a matrix factorization (MF) \cite{koren2009matrix} model with interaction data for recommendation in the target domain.

\subsubsection{Global-Global Contrast}
 Global-level contrast is often used in sequential recommendation models where a sequence is considered a user's global view, which can be formulated as:
\begin{equation}
    f_{\theta}^{*}=\mathop{\arg\min}\limits_{f_{\theta}, g_{\phi_{s}}}\mathcal{L}_{\mathcal{MI}}\big(g_{\phi_{s}}(\text{Agg}(f_{\theta}(\tilde{S}_{i})),\text{Agg}(f_{\theta}(\tilde{S}_{j}))\big),     
\end{equation}
where $\tilde{S}_{i}$ and $\tilde{S}_{j}$ are two sequence augmentations, and $\text{Agg}$ is the aggregating function that synthesizes sequence representations based on involved item representations.

As a representative model, \textbf{CL4SRec} \cite{xie2020contrastive} uses three random augmentation methods - item masking, item cropping, and item reordering - to augment sequences. It applies these methods to $N$ sequences, resulting in 2$N$ augmented sequences  $\left[S_{u_{1}}^{\prime}, S_{u_{1}}^{\prime\prime}, S_{u_{2}}^{\prime}, S_{u_{2}}^{\prime\prime}, \ldots, S_{u_{N}}^{\prime}, S_{u_{N}}^{\prime\prime}\right]$. CL4SRec treats the pairs $(S_{u}^{\prime}, S_{u}^{\prime\prime})$ as positive pairs and uses 2($N$-1) augmented sequences as negative samples within the mini-batch. A Transformer-based encoder \cite{vaswani2017attention} is then used to encode the augmented sequences and learn user representations for global-level contrast. This approach is also adopted in \textbf{H$^{2}$SeqRec} \cite{li2021hyperbolic}, \textbf{CoSeRec} \cite{liu2021contrastive}, \textbf{ContraRec} \cite{wangsequential}, and \textbf{UniSRec} with some variations. CoSeRec substitutes items in short sequences with correlated items or inserts correlated items into short sequences for robust data augmentation, while ContraRec not only contrasts sequences augmented from the same input but also considers sequences with the same target item as positive pairs. UniSRec further proposes to contrast sequence representations from different domains for universal representation learning. Besides, \textbf{DHCN} \cite{xia2021aaai} creates two hypergraph-based views of a given session by modeling the intra-session and inter-session structural information. It regards the representations of the same session as positive pairs and the corrupted representations (obtained by feature shuffling) of different sessions as negatives.

\subsubsection{Local-Global Contrast} 
Local-global contrast aims to encode high-level global information into local structure representations and unify global and local semantics. It is frequently used in the graph learning scenario, where it can be formulated as:
\begin{equation}
    f_{\theta}^{*}=\mathop{\arg\min}\limits_{f_{\theta}, g_{\phi_{s}}}\mathcal{L}_{\mathcal{MI}}\big(g_{\phi_{s}}(\tilde{\mathbf{h}},\mathcal{R}(f_{\theta}(\tilde{\mathcal{G}},\tilde{\mathbf{A}}))\big),     
\end{equation}
where $\mathcal{R}$ is the readout function that generates global-level graph representation.

\textbf{EGLN} \cite{yang2021enhanced} proposes to achieve local-global consistency by contrasting merged user-item pair representations with the global representation, which is an average of all the user-item pair representations. It also adopts graph diffusion for data augmentation and obtains an augmented graph adjacency matrix by calculating the similarities between users and items, retaining the top-$K$ similarities. The matrix and the user/item representations iteratively learn from each other and get updated via a graph encoder. Similarly, \textbf{BiGI} \cite{cao2021bipartite} performs local-global contrast, but only its $h$-hop subgraph is sampled for feature aggregation when generating the user-item pair representation. In \textbf{HGCL} \cite{cai2022heterogeneous}, user and item node-type specific homogeneous graphs are constructed. For each homogeneous graph, it maximizes the mutual information between local patches of a graph and the global representation of the entire graph using DGI \cite{velickovic2019deep} pipeline. It also proposes a cross-type contrast to measure local and global information across different types of homogeneous graphs.

\subsubsection{Local-Context Contrast}
Local-context contrast is observed in both graph and sequence-based scenarios, where context is constructed by sampling ego-networks or clustering. This type of contrast can be formulated as:
\begin{equation}
    f_{\theta}^{*}=\mathop{\arg\min}\limits_{f_{\theta}, g_{\phi_{s}}}\mathcal{L}_{\mathcal{MI}}\big(g_{\phi_{s}}(\mathbf{h}_{i},\mathcal{R}(f_{\theta}(\mathcal{C}_{j}))\big),     
\end{equation}
where $\mathcal{C}_{j}$ denotes the context of node (sequence) $j$.

Inspired by \cite{li2020prototypical}, \textbf{NCL} \cite{lin2022improving} designs a prototypical contrastive objective to capture the correlations between a user/item and its prototype, which represents a group of semantic neighbors. The prototype is obtained by clustering over all user or item embeddings using the K-means algorithm, and the EM algorithm is used to adjust the prototypes recursively. \textbf{ICL} \cite{chen2022intent} has a similar pipeline, but it is designed for sequential recommendation where the semantic prototypes are modeled as user intents and the belonged sequence is a local view of the prototype. \textbf{MHCN} \cite{yu2021self} applies SSL to social recommendation \cite{yin2016spatio} by defining three types of triangle social relations and modeling them with a multi-channel hypergraph encoder. For each user in each channel, MHCN hierarchically maximizes the mutual information among the user representation, the user's ego hypergraph representation, and the global hypergraph representation. Compared to MHCN, \textbf{HCCF} \cite{xia2022hypergraph} proposes to parameterize hypergraph dependency matrices instead of manually defining hypergraph structures. It then contrasts  representations derived from the user-item graph and the parameterized hypergraph.  \textbf{SMIN} \cite{long2021social} contrasts nodes with their contexts using a chain of user-item adjacency matrices with different orders for context aggregation. \textbf{S$^{3}$-Rec} \cite{zhou2020s3} applies item masking and item cropping to augment sequences and devises four contrastive tasks for pre-training a bidirectional Transformer for next-item prediction: item-attribute mutual information maximization (MIM), sequence-item MIM, sequence-attribute MIM, and sequence-sequence MIM.

\subsection{Feature-Level Contrast}
Compared to structure-level contrast, feature-level contrast is relatively less explored due to limited feature/attribute information in academic datasets. However, in industry, data is often organized in a multi-field format, and a large number of categorical features such as user profile and item category can be used. Generally, this type of contrast can be formally defined as:
\begin{equation}
	f_{\theta}^{*}=\mathop{\arg\min}\limits_{f_{\theta}, g_{\phi_{s}}}\mathcal{L}_{\mathcal{MI}}\big(g_{\phi_{s}}(f_{\theta}(\tilde{\mathbf{x}}_{i}),f_{\theta}(\tilde{\mathbf{x}}_{j}))\big),     
\end{equation}
where $\tilde{\mathbf{x}}_{i}$ and $\tilde{\mathbf{x}}_{j}$ are feature-level augmentations which are obtained by modifying input feature or learned by models.  

\textbf{SL4Rec} \cite{yao2021self} adopts a two-tower architecture and applies correlated feature masking and dropout on the item features for more meaningful feature augmentations. It seeks to mask highly correlated features together, measured by mutual information, but this makes the contrastive task challenging because retained features may not remedy the semantics behind the masked features. \textbf{SLMRec} \cite{tao2022self} considers the multi-modalities of recommendation data and leverages feature dropout and masking to create data augmentations with different granularities. It then conducts a modal-agnostic contrastive task and a modal-specific contrastive task to distill additional supervision signals lying in different modalities. \textbf{MISS} \cite{guo2021miss} argues that directly perturbing a user behavior sequence may cause semantically dissimilar augmentations since a sequence may contain multiple interests. Instead, it uses a CNN-based multi-interest extractor to transform user samples, containing behavior data and categorical features, into a group of implicit interest representations that augment the user sample at the feature-level. The contrastive task is then conducted on the extracted interest representations.

\subsection{Model-Level Contrast}
The former two categories extract self-supervision signals from the data perspective, but they are not implemented in an end-to-end fashion. Another way is to keep the input unchanged and dynamically modify the model architecture to augment view pairs on-the-fly. The contrast between these model-level augmentations is formulated as:
\begin{equation}
	f_{\theta}^{*}=\mathop{\arg\min}\limits_{f_{\theta}, g_{\phi_{s}}}\mathcal{L}_{\mathcal{MI}}\big(g_{\phi_{s}}(f_{\theta^{\prime}}(\mathcal{D}),f_{\theta^{\prime\prime}}(\mathcal{D}))\big),     
\end{equation}
where $f_{\theta^{\prime}}$ and $f_{\theta^{\prime\prime}}$ are perturbed versions of $f_{\theta}$. This equation can be seen as a special case of Eq. (\ref{eq: contrastive}) which augments the intermediate hidden representations of the same input.

Neuron masking is a common technique used to perturb the model. \textbf{DuoRec} \cite{Qiuwsdm22} applies different dropout masks to a Transformer-based backbone for two model-level representation augmentations, maximizing the mutual information between the two representations. This method appears simple but shows significant performance in next-item prediction tasks. \textbf{SimGCL} \cite{yu2021graph} and \textbf{XSimGCL} \cite{yu2022xsimgcl} add random uniform noise to hidden representations for augmentations, resulting in more uniform node representations that mitigate the popularity bias issue \cite{chen2020bias}. Adjusting the noise magnitude can provide finer-grained regulation of representation uniformity, leading to advantages over SGL on recommendation accuracy and model training efficiency. In particular, XSimGCL further proposes the cross-layer contrast. In \textbf{SRMA} \cite{liu2021modelaug}, apart from neuron-level perturbations, it proposes randomly dropping some layers of the feed-forward network in the Transformer for model-level augmentations. Additionally, it introduces another pre-trained encoder with the same architecture but trained with the recommendation task to generate views for contrast.

% Beyond the taxonomy, there are a few papers \cite{wei2022contrastive,wu2022multi} which claim that their proposed methods are contrastive self-supervised. They deal with multi-behavior data and use auxiliary behavior data as supervision signals for data augmentation. Views of the same user's behavior are considered positive pairs, while views of different users are sampled as negative pairs, and the methods encourage consistency between representations of positive pairs by conducting behavior-level contrast. However, we think it is farfetched to call them self-supervised because they do not transform the original data to create new views.

\begin{table*}[t]
    \caption{A summary of the surveyed papers on contrastive self-supervised recommendation.}
	\label{Table: contrastive methods}
    \scriptsize
    \renewcommand\arraystretch{1.0}
    \begin{center}
        \resizebox{\textwidth}{!}{
    \begin{tabular}{lccccc}
    \toprule
    \textbf{Method} & \textbf{Scenario}  & \textbf{Data Augmentation} &   \textbf{Contrast Type}  & \textbf{Contrastive Objective} & \textbf{Training Scheme} \\  \hline
    SGL\cite{wujc2021self}    & Graph                        & Edge/Node Dropout & Node-Node & InfoNCE & Joint Learning              \\  \hline
    DCL \cite{liu2021contrastiveL}   & Graph                        & Edge Dropout     & Node-Node & InfoNCE & Joint Learning              \\ \hline  
    CCDR \cite{xie2021contrastive}  & Graph (Cross-domain)         & Subgraph Sampling   & Node-Node & InfoNCE & Joint Learning              \\  \hline
    PCRec \cite{wang2021pre} & Graph (Cross-domain)         & Subgraph Sampling      & Node-Node & InfoNCE & Pre-training\&Fine-tuning \\ \hline  
    HHGR \cite{zhang2021double}  & Graph (Group) & Node Dropout        & User-User  & Cross-Entropy & Joint Learning        \\  \hline
    KGCL \cite{yang2022knowledge}  & Graph (Knowledge) & Node/Edge Dropout        & Node-Node  & InfoNCE & Joint Learning        \\  \hline          
    CL4SRec \cite{xie2020contrastive}   &  Sequential       & Item Masking/Reordering/Cropping     & Sequence-Sequence & InfoNCE & Joint Learning\\ \hline
    H$^{2}$SeqRec \cite{li2021hyperbolic}   &  Sequential       & Item Masking/Cropping  & Sequence-Sequence & InfoNCE & Pre-training\&Fine-tuning\\ 
    \hline
    CoSeRec \cite{liu2021contrastive}   &  Sequential       & Item Substitution/Insertion     & Sequence-Sequence & InfoNCE & Joint Learning\\ \hline
    ContraRec \cite{wangsequential}  &  Sequential       & Item Masking/Reordering/overlapping     & Sequence-Sequence & InfoNCE & Joint Learning\\ \hline            
    DHCN \cite{xia2021aaai}   &  Session       & Feature Shuffling     & Session-Session & Cross-Entropy & Joint Learning\\ \hline  
    UniSRec \cite{hou2022towards}   &  Sequential       & item/word dropout     & Sequence-sequence/item & InfoNCE & Pre-training\&Fine-tuning\\ \hline   
    EGLN \cite{yang2021enhanced}   &  Graph       & Graph Diffusion     & Pair-Graph & Cross-Entropy & Joint Learning\\ \hline   
    BiGI \cite{cao2021bipartite}   &  Graph       & Subgraph Sampling     & Pair-Graph & Cross-Entropy & Joint Learning\\ \hline
    HGCL \cite{cai2022heterogeneous} & Graph      & Feature Shuffling    & Node-Graph & Cross-Entropy & Joint Learning \\ \hline   
    MHCN \cite{yu2021self}   &  Graph (Social)    & Subgraph Sampling/Feature Shuffling     & User-Hypergraph & Triplet-loss & Joint Learning\\ \hline
    HCCF \cite{xia2022hypergraph}   &  Graph    & Edge Dropout    & User-Hypergraph & InfoNCE & Joint Learning\\ \hline
    SMIN \cite{long2021social}   &  Graph (Social)    & Graph Diffusion     & Node-Context & Cross-Entropy & Joint Learning\\ \hline
    NCL \cite{lin2022improving}   &  Graph    & Feature Clustering     & Node-Cluster & InfoNCE & Joint Learning\\ \hline
    ICL \cite{chen2022intent}   &  Sequential    & Feature Clustering     & Sequence-Cluster & InfoNCE & Joint Learning \\\hline   
    S$^{3}$-Rec \cite{zhou2020s3}  &  Sequential    & Item Masking/Cropping    & Item-Context & InfoNCE & Pre-training\&Fine-tuning\\ \hline
    SL4Rec \cite{yao2021self}   &  Feature-Based      & Feature Dropout     & Item Feature-Item Feature & InfoNCE & Joint Learning\\ \hline
    SLMRec \cite{tao2022self} &  Graph       & Feature Dropout/Masking     & Modality-Modality & InfoNCE & Joint Learning\\ \hline
    MISS \cite{guo2021miss}   &  CTR Prediction    & Feature Extractor     & User Feature-User Feature & InfoNCE & Joint Learning\\ \hline
    DuoRec \cite{Qiuwsdm22}  &  Sequential       & Neuron Masking     & Sequence-Sequence & InfoNCE & Joint Learning\\ \hline
    SimGCL \cite{yu2021graph}   &  Graph           & Feature Noises     & Node-Node & InfoNCE & Joint Learning\\  \hline 
    XSimGCL \cite{yu2022xsimgcl}   &  Graph           & Feature Noises     & Node-Node & InfoNCE & Joint Learning\\  \hline    
    SRMA \cite{liu2021modelaug}  &  Sequential       & \makecell{Neuron Masking/Layer Dropping\\/Encoder Complementing}    & Sequence-Sequence & InfoNCE & Joint Learning\\     
    \bottomrule
\end{tabular}
        }
\end{center}
\end{table*}

\subsection{Contrastive Loss} \label{sec:cl loss}
The contrastive loss is a research hotspot in the machine learning community \cite{wang2020understanding,wang2021understanding}, and is also drawing increasing attention in SSR. Generally, the optimization goal of the contrastive loss is to maximize the mutual information (MI) between two representations $\mathbf{h}_{i}$ and $\mathbf{h}_{j}$ defined as:
\begin{equation}
    \begin{aligned}
    \mathcal{MI}\left(\mathbf{h}_{i}, \mathbf{h}_{j}\right) =\mathbb{E}_{P\left(\mathbf{h}_{i}, \mathbf{h}_{j}\right)}\log \frac{P\left(\mathbf{h}_{i}, \mathbf{h}_{j}\right)}{P\left(\mathbf{h}_{i}\right) P\left(\mathbf{h}_{j}\right)}
    \end{aligned}
\end{equation}
However, directly maximizing MI is difficult and a practical way is to maximize its lower bound. In this section, we review two most commonly used lower bounds: Jensen-Shannon Estimator \cite{nowozin2016f} and InfoNCE \cite{oord2018representation}. 

\subsubsection{Jensen-Shannon Estimator}
As a MI estimator for SSL, Jensen-Shannon divergence (JSD) first appears in DGI \cite{velickovic2019deep}. Compared to the Donsker-Varadhan estimator \cite{belghazi2018mutual} which provides a tight lower bound of MI, JSD can be more efficiently optimized and guarantees a stable performance (we refer you to \cite{hjelm2018learning} for a detailed derivative of how JSD estimates MI). It is widely used in the graph scenario, and can be formulated as:     
\begin{equation}
    \begin{aligned}
    \mathcal{MI}_{\mathrm{JSD}}\left(\mathbf{h}_{i}, \mathbf{h}_{j}\right) &=-\mathbb{E}_{\mathcal{P}}\left[\log \left(f_{\mathrm{D}}\left(\mathbf{h}_{i}, \mathbf{h}_{j}\right)\right)\right] \\
    &-\mathbb{E}_{\mathcal{P} \times \tilde{\mathcal{P}}}\left[\log \left(1-f_{\mathrm{D}}\left(\mathbf{h}_{i}, \tilde{\mathbf{h}}_{j}\right)\right)\right]
    \end{aligned}
\end{equation}
where $\mathcal{P}$ denotes the joint distribution of $\mathbf{h}_{i}$ and $\mathbf{h}_{j}$, and $\mathcal{P} \times \tilde{\mathcal{P}}$ denotes the product of marginal distributions. The discriminator $f_{\mathrm{D}}: \mathbb{R}^{d} \times \mathbb{R}^{d} \rightarrow \mathbb{R}$ can be implemented in various forms. The vanilla implementation in \cite{velickovic2019deep} is $f_{\mathrm{D}} = \mathbf{h}_{i}^{\top}\mathbf{W}\mathbf{h}_{j}$ which is called bilinear scoring function and is directly applied in \cite{yang2021enhanced,cao2021bipartite}. In \cite{xia2021aaai,long2021social}, its dot-product form is used and shows comparable performance.

\subsubsection{Noise-Contrastive Estimator}
InfoNCE \cite{oord2018representation} uses a softmax-based version of NCE \cite{gutmann2010noise} to identify positive samples among a set of negative samples. It is shown that InfoNCE often outperforms JSD in CV tasks \cite{hjelm2018learning}. As the most popular contrastive loss in SSR, InfoNCE is formulated as:
\begin{equation}
\label{eq:infonce}
\mathcal{MI}_{\mathrm{NCE}}=-\mathbb{E}\bigg[\log \frac{e^{f_{\mathrm{D}}(\mathbf{h}_{i}, \mathbf{h}_{j})}}{\sum_{n \in \mathcal{N}_{i}^{-}\cup\{j\}}e^{f_{\mathrm{D}}(\mathbf{h}_{i}, \mathbf{h}_{n})}}\bigg]
\end{equation}  
where $\mathcal{N}_{i}^{-}$ is the negative sample set of $i$, and is often sampled within a batch. When $f_{\mathrm{D}}(\mathbf{h}_{i}, \mathbf{h}_{j})=\frac{\mathbf{h}_{i}^{\top}\mathbf{h}_{j}}{\|\mathbf{h}_{i}\|\|\mathbf{h}_{j}\|}/\tau$, it is also called NT-Xent loss \cite{chen2020simple} where $\tau$ is the temperature (e.g., 0.2), which is its well-known version.

Despite the effectiveness of InfoNCE, its behaviors in SSR has not been well investigated. Wang and Isola \cite{wang2020understanding} identify two key properties related to its NT-Xent version: \textit{alignment} (closeness) of representations from positive pairs and \textit{uniformity} of the normalized representations on the hypersphere. Rewriting the NT-Xent loss, we derive:
\begin{equation}
    \underbrace{\mathbb{E}\left[-\mathbf{z}_{i}^{\top}\mathbf{z}_{j} / \tau\right]}_{\text {alignment }}+\underbrace{\mathbb{E}\left[\log \left(e^{\mathbf{z}_{i}^{\top}\mathbf{z}_{j} / \tau}+\underset{n \in \mathcal{N}_{i}^{-}}{\sum} e^{\mathbf{z}_{i}^{\top}\mathbf{z}_{n} / \tau}\right)\right]}_{\text {uniformity }}
\end{equation}
where $\mathbf{z}_{i}=\frac{\mathbf{h}_{i}}{\|\mathbf{h}_{i}\|}$. In SSR, a few studies have reported their existence. Qiu \textit{et al.} \cite{Qiuwsdm22} and Yu \textit{et al.} \cite{yu2021graph} demonstrate that optimizing the InfoNCE loss learns a more even distribution of item/node representations, which can mitigate the representation degeneration problem and address the popularity bias \cite{wu2022effectiveness}. In addition, Wang and Liu \cite{wang2021understanding} show that the InfoNCE loss is hardness-aware, and the temperature $\tau$ controls the strength of penalties on hard negative samples. 

Meanwhile, there is also a potential downside of the InfoNCE loss. For each instance in the input, InfoNCE pushes it away from other instances except its augmentation counterpart in the representation space. However, the existence of similar users/items leads to a large number of false negative samples, which will impair recommendation performance. To tackle this problem, a few works \cite{Qiuwsdm22,yu2021kdd,xia2021cotrec} propose to incorporate multiple positive samples into the InfoNCE loss, deriving a modified version:
\begin{equation}
    \mathcal{MI}^{+}_{NCE}=-\mathbb{E}\bigg[\log \frac{\sum_{j \in \mathcal{N}_{i}^{+}}e^{-\mathbf{z}_{i}^{\top}\mathbf{z}_{j} / \tau}}{\sum_{n \in \mathcal{N}_{i}^{-}\cup\mathcal{N}_{i}^{+}}e^{-\mathbf{z}_{i}^{\top}\mathbf{z}_{n} / \tau}}\bigg].
\end{equation}  
Qiu \textit{et al.} \cite{Qiuwsdm22} proposes to identify semantically positive sequences by checking if two sequences have the same item to be predicted. Yu \textit{et al.} \cite{yu2021kdd} and Xia \textit{et al.} \cite{xia2021cotrec} build multiple encoders on semantically similar graphs to predict positive examples for a given instance.

\subsection{Pros and Cons}
Due to the flexibility to augment data and set pretext tasks, contrastive methods expand rapidly in recent years and reach out most recommendation topics. While contrastive SSR has shown remarkable effectiveness in improving recommendation with lightweight architectures, it is often compromised by \textbf{the unknown criterion for high-quality data augmentations} \cite{tian2020makes}. Existing contrastive methods are mostly based on arbitrary data augmentations and are selected by trial-and-error. There have been neither rigorous understanding of how and why they work nor rules or guidelines clearly telling what good augmentations are for recommendation. In addition, some common augmentations, which were considered useful, recently even have been proved having a negative impact on recommendation performance \cite{yu2021graph}. As a result, without knowing what augmentations are informative, the contrastive task may fail.

	\section{Generative Methods}\label{sec:generative}
Generative SSR methods aim to encode intrinsic correlations in data by reconstructing the original input with its corrupted version. We divide these methods into two categories based on their reconstruction objectives: \textbf{Structure Generation} and \textbf{Feature Generation}. The surveyed generative methods are summarized in Table \ref{Table: generative methods}
\subsection{Structure Generation}
This branch of methods leverages the structural information to supervise the model. By applying the masking/dropout-based augmentation operators (see Section \ref{sec:augmentation}) to the original structure, its corrupted versions are obtained. In the scenario of sequence-based recommendation, recovering the structure can be formulated as:
\begin{equation}
	f_{\theta}^{*}=\underset{f_{\theta}, g_{\phi_{s}}}{\arg \min } \mathcal{L}_{ssl}\left(g_{\phi_{s}}\left(f_{\theta}(\tilde{\mathcal{S}})\right), S\right)
\end{equation}
where $\tilde{\mathcal{S}}$ denotes the corrupted sequence in which a portion of items are masked (replaced with a special token [$\mathrm{mask}$]). 

\textbf{BERT4Rec} \cite{sun2019bert4rec} is the first to use BERT \cite{devlin2018bert} for sequential recommendation, with \textbf{BERT4SessRec} \cite{chen2019bert4sessrec} being a similar concurrent work. It enhances the left-to-right training approach in SASRec \cite{kang2018self} by learning a bidirectional representation model. The method randomly masks some items in input sequences and predicts the masked items based on surrounding items. To indicate the item to be predicted, it appends the [$\mathrm{mask}$] token at the end of the input sequence. The objective is formulated as:
\begin{equation}
	\mathcal{L}=\frac{1}{\left|S_{u}^{m}\right|} \sum_{i_{m} \in S_{u}^{m}}-\log P\left(i_{m}=i_{m}^{*} \mid \tilde{S}_{u}\right),
\end{equation}
where $\tilde{S}_{u}$ is the corrupted version of $S_{u}$, $S_{u}^{m}$ is the randomly masked items, and $i^{*}_{m}$ is the true masked item. This loss is widely used in the BERT-like SSR models, serving as the optimization objective of their generative pretext tasks. 

Inspired by the success of BERT4Rec, follow-up works apply the masked-item-prediction training to more specific scenarios. For instance, \textbf{UNBERT} \cite{zhang2021unbert} and Wu \textit{et al.} \cite{wu2021empowering} have explored the technique for news recommendation in similar ways. The input of UNBERT is a combination of news sentences and user sentences with a set of special symbols. It randomly masks some word-piece tokens to pre-train the token representations with the Cloze task, and then fine-tunes the model on the news recommendation task. \textbf{U-BERT} \cite{qiu2021u} uses review comments to pre-train the encoder with masked-word-token-prediction in the source domain, and then fine-tunes the encoder with an added layer in the target domain where comments are insufficient for rating prediction. In addition, \textbf{GRec} \cite{yuan2020future} develops a gap-filling mechanism with the encoder-decoder setting. The encoder takes a partially-complete session sequence as input, and the decoder predicts the masked items conditioned on the both the output of the encoder and its own complete embeddings. \textbf{UPRec} \cite{xiao2021uprec} further modifies BERT4Rec to enable it to exploit heterogeneous information such as user attributes and social networks to enhance the sequence modeling.

\begin{table*}[t]
    \caption{A summary of the surveyed papers on generative self-supervised recommendation.}
	\label{Table: generative methods}
    \renewcommand\arraystretch{1.0}
    \scriptsize
    \begin{center}
		\resizebox{\textwidth}{!}{
    \begin{tabular}{lcccc}
    \toprule
    \textbf{Method} & \textbf{Scenario}  & \textbf{Data Augmentation} &   \textbf{Branch} & \textbf{Training Scheme} \\   \hline
    BERT4Rec \cite{sun2019bert4rec}    & Sequential      &Item Masking &  Structure Generation & Integrated Learning              \\  \hline
	BERT4SessRec \cite{chen2019bert4sessrec} & Sequential (Session)      &Item Masking &  Structure Generation & Pre-training\&Fine-tuning              \\  \hline
	UNBERT \cite{zhang2021unbert}    & Sequential (News) &Word Masking &  Structure Generation & Pre-training\&Fine-tuning              \\  \hline
	U-BERT \cite{qiu2021u}    & Sequential      &Word Masking &  Structure Generation & Pre-training\&Fine-tuning              \\  \hline
	GRec \cite{yuan2020future}    & Sequential (Session)      &Item Masking &  Structure Generation & Integrated Learning              \\  \hline
	UPRec \cite{xiao2021uprec}    & Sequential      &Item Masking &  Structure Generation & Pre-training\&Fine-tuning              \\  \hline
	PeterRec \cite{yuan2020parameter}    & Sequential      &Item Masking &  Structure Generation & Pre-training\&Fine-tuning              \\  \hline
	ShopperBERT \cite{shin2021one4all}    & Sequential      &Item Masking &  Structure Generation & Pre-training\&Fine-tuning              \\  \hline
	P5 \cite{geng2022recommendation} & Multiple Tasks    &Token Masking &  Structure Generation & Pre-training\&Prompt-tuning              \\  \hline
	G-BERT \cite{shang2019pre}    & Graph (Medication)      &Node Masking &  Structure Generation & Pre-training\&Fine-tuning              \\  \hline
	PMGT \cite{liu2021pre}    & Graph      &Subgraph Sampling/Node Masking &  Structure \& Feature Generation & Pre-training\&Fine-tuning              \\  \hline
	Ma \textit{et al.} \cite{ma2020disentangled} & sequential  & Sequence Splitting & Feature Generation & Joint Learning \\ \hline
	MMInfoRec \cite{qiu2021memory} & sequential  & Sequence Splitting/Feature Dropout & Feature Generation & Integrated Learning \\ \hline
	PT-GNN \cite{hao2021pre}   & Graph    &Subgraph Sampling &  Feature Generation & Pre-training\&Fine-tuning              \\  
    \bottomrule
\end{tabular}
		}
\end{center}
\end{table*}

The generative models discussed above are mainly pre-trained for a specific recommendation task. However, there is also a line of research that aims to learn general-purpose representations through generative pre-training, benefiting multiple downstream recommendation tasks. \textbf{PeterRec} \cite{yuan2020parameter} and \textbf{ShopperBERT} \cite{shin2021one4all} are two examples. PeterRec transfers the pre-trained model parameters to user-related tasks by injecting small grafting neural networks into the original model and training only these patches to adapt to specific tasks. Similarly, \textbf{ShopperBERT} \cite{shin2021one4all} is pre-trained with nine generative pretext tasks including the masked-purchase-prediction, and the learned universal user representations can serve six downstream recommendation-related tasks and shows superiority over the task-specific Transformer-based models which are learned from scratch. A very recent work \textbf{P5} \cite{geng2022recommendation} takes the first step to explore the combination of prompt learning \cite{liu2023pre} and personalization. It converts all recommendation data into natural language sequences and learns different tasks with the same language modeling objective during pre-training, making it a foundation model for various downstream recommendation tasks.

In the graph-based recommendation scenario, structure generation is formulated as:
\begin{equation}
	f_{\theta}^{*}=\underset{f_{\theta}, g_{\phi_{s}}}{\arg \min } \mathcal{L}_{ssl}\left(g_{\phi_{s}}\left(f_{\theta}(\tilde{\mathcal{G}})\right), \mathbf{A}\right)
\end{equation}

\textbf{G-BERT} \cite{shang2019pre} combines GNNs and BERT for medication recommendation. It models the diagnosis and medication codes in electrical health records as two tree-like graphs and employs GNNs to learn the graph representations, which are then fed to a BERT encoder and pre-trained with two generative pretext tasks: self-prediction and dual-prediction. The self-prediction task reconstructs masked codes with the same type of graph, while the dual-prediction task reconstructs masked codes with the other type of graph. \textbf{PMGT} \cite{liu2021pre} conducts a graph reconstruction task with the sampled subgraph. It develops a sampling method to sample subgraph for each item node, and reorganizes the sampled subgraph as an ordered sequence according to the neighbor importance. The subgraphs are then fed to a Transformer encoder, and the method pre-trains the item representations with the missing neighboring item prediction. 

\subsection{Feature Generation}
Feature generation can be understood as a regression problem and formulated as:
\begin{equation}
	f_{\theta}^{*}=\underset{f_{\theta}, g_{\phi_{s}}}{\arg \min }\left\|g_{\phi_{s}}\left(f_{\theta}(\tilde{\mathcal{D}})\right)-\mathbf{X}\right\|^{2},
\end{equation}
where $\|\cdot\|^{2}$ is the MSE loss, and $\mathbf{X}$ is a general symbol of features that can be the user profile attributes, item textual features, or learned user/item representations.

Inspired by GPT-GNN \cite{hu2020gpt}, \textbf{PMGT} \cite{liu2021pre} also sets up a feature reconstruction task to pre-tain the Transformer-based recommendation model. The method initializes item embeddings with the extracted image and textual features and then masks a portion of sampled nodes and uses the remaining nodes to recover the features of the masked ones. As for the sequence feature generation, Ma \textit{et al.} \cite{ma2020disentangled} propose to reconstruct the representation of the future sequence with the past behaviors. Specifically, they disentangle the intentions behind any given sequence of behaviors and the reconstruction is conducted between any pairs of sub-sequences that involve a shared intention. Similarly, in \textbf{MMInfoRec} \cite{qiu2021memory}, given a sequence with $t$ items, it encodes the sequence and predicts the next item's representation at time step $t$+1. The augmented sequence representation is then compared with the ground-truth. An auto-regressive prediction module is designed to include more futuristic information by predicting the $t$+$k$ item with item $t$+1 to item $t$+$i$-1, with a dropout function used to create multiple semantically similar item representations. To enhance cold-start users and items representation, \textbf{PT-GNN} \cite{hao2021pre} proposes to pre-train GNN models by mimicking the meta-learning setting. It picks the users/items with sufficient interactions as target users/items, and performs graph convolution on sampled $K$ neighbors for the targets to predict their ground-truth embeddings learned from the whole graph. Optimizing this reconstruction loss directly improves the embedding capacity, making the model easily and rapidly adapt to cold-start users/items. 

\subsection{Pros and Cons}
The latest generative SSR methods generally follow the masked language model pipeline and rely on the power of Transformers to achieve significant results. The successful training of BERT on large-scale datasets has opened the way for large MLM-based recommendation models. However, these methods may face the challenge of \textbf{intensive computation}. Most current Transformer-based generative methods are trained on small datasets typically and integrate only one or two blocks. However, training with large-scale datasets for news recommendation or general-purpose representations can be extremely computationally demanding. Considering that scaling-up pre-training datasets to build powerful large models has become the trend across different AI communities, it truly torments research groups with limited computing resources.

	\section{Predictive Methods}\label{sec:predictive}
Predictive SSR methods deal with self-generated supervisory signals obtained from the complete original data. According to what the predictive self-supervised task predicts, we divide predictive methods into two branches: \textbf{Sample Prediction} and \textbf{Pseudo-Labels Prediction}. A summary of the surveyed predictive methods can be found in Table \ref{Table: predictive methods}

\subsection{Sample Prediction}
Self-training \cite{zoph2020rethinking}, a flavor of semi-supervised learning, is linked to SSL in the Sample Prediction branch. The SSR model is pre-trained on the original data, and potential informative samples for the recommendation task are predicted using the pre-trained parameters as augmented data. These samples are then used to enhance the recommendation task or recursively generate better samples. The difference between SSL-based sample prediction and pure self-training is that in semi-supervised learning, a finite number of unlabeled samples are available, while in SSL, samples are dynamically generated.

Sequential recommendation models often perform poorly on short sequences due to limited user behaviors. To improve the model performance, \textbf{ASReP} \cite{liu2021augmenting} proposes to augment the short sequences with pseudo-prior items. Given ordered sequences, ASRep first pre-trains a Transformer-based encoder SASRec \cite{kang2018self} in a reverse manner (i.e., from right-to-left) so that the encoder is capable of predicting the pseudo-prior items. An augmented sequence is obtained by appending the fabricated subsequence to the beginning of the original sequence. The encoder is then fine-tuned on the augmented sequences in a left-to-right manner to predict the next item in the original sequence. A follow-up work \textbf{BiCAT} \cite{jiang2021sequential} argues that the reverse augmentation may be inconsistent with the original correlation. It further proposes to simultaneously pre-train the encoder from both left-to-right and right-to-left directions. The bidirectional training can bridge the gap between the reverse augmentation and the forward recommendation. In the graph scenario, the samples can also be predicted based on node feature/semantic similarities. When there are multiple encoders built on different graphs, they can recursively predict samples for other encoders where the self-training is upgraded to co-training \cite{blum1998combining}. We will find this idea in \textbf{SEPT} \cite{yu2021kdd} and \textbf{COTREC} \cite{xia2021cotrec} which are introduced in Section \ref{sec:Hybrid}. 

\begin{table*}[t]
    \caption{A summary of the surveyed papers on predictive self-supervised recommendation.}
	\label{Table: predictive methods}
    \renewcommand\arraystretch{1.0}
    \scriptsize
    \begin{center}
        \resizebox{\textwidth}{!}{
    \begin{tabular}{lp{2cm}<{\centering}p{6cm}<{\centering}cc}
    \toprule
    \textbf{Method} & \textbf{Scenario}  & \textbf{Data Augmentation} &   \textbf{Branch} & \textbf{Training Scheme} \\  \hline
    ASReP \cite{liu2021augmenting}    & Sequential      &Prior Items &  Sample Prediction & Pre-training\&Fine-tuning              \\  \hline
    BiCAT \cite{jiang2021sequential}     & Sequential      &Prior Items &  Sample Prediction & Pre-training\&Fine-tuning              \\  \hline
    PTUM \cite{WuWQLH020}   & Sequential  &  Matching Label &  Relation Prediction & Pre-training\&Fine-tuning \\ \hline
    SSI \cite{YuanCSZD21}   & Sequential  &  Matching Label &  Relation Prediction & Pre-training\&Fine-tuning \\ \hline
    CHEST \cite{wang2021curriculum}   & Graph &   Path Label &  Relation Prediction & Pre-training\&Fine-tuning \\ \hline      
    BUIR \cite{lee2021bootstrapping}   & Graph &  Bootstrapped representation &  Similarity Prediction & Integrated Learning \\ \hline
    SELFCF \cite{zhou2021selfcf}   & Graph &  Bootstrapped representation &  Similarity Prediction & Integrated Learning \\ \hline
    CLUE \cite{cheng2021learning}  & Sequential &  Bootstrapped representation &  Similarity Prediction & Integrated Learning \\ \hline
    RDC \cite{liuhc2021self} &Graph& Learned Rating Distribution & Similarity Prediction & Joint Learning \\ \hline
    MrTransformer \cite{ma2021improving} & Sequential & Recombined Preference Feature & Similarity Prediction & Joint Learning \\ \hline
    DUAL \cite{tao2022predictive} & Graph (Social) & Pre-computed Probability & Similarity Prediction & Joint Learning \\
    \bottomrule
\end{tabular}
}
\end{center}
\end{table*}

\subsection{Pseudo-Label Prediction}
In this branch, there are two forms of pseudo-labels: pre-defined discrete values and pre-computed/learned continuous values. The former describes a relation between two objects, and the corresponding predictive task predicts whether the relation exists. The latter describes an attribute value of the given object (e.g. node degree), a probability distribution, or a feature vector. The corresponding predictive task aims to minimize the difference between the output and the pre-computed continuous values. These prediction tasks can be formulated as \textbf{Relation Prediction} and \textbf{Similarity Prediction}.

\subsubsection{Relation Prediction}
The relation prediction task can be formulated as a classification problem where the pre-defined relations, serving as the pseudo-labels, are self-generated at no cost. We can refine Eq. (\ref{eq: predictive}) to: 
\begin{equation}
    f_{\theta^{*}}=\underset{f_{\theta}, g_{\phi_{s}}}{\arg \min } \mathcal{L}_{ce}\left(g_{\phi_{s}}\left(f_{\theta}(o_{i},o_{j})\right), \mathcal{T}(o_{i},o_{j})\right),
\end{equation}
where $o_{i}$ and $o_{j}$ are a pair of objects from $\mathcal{D}$, $\mathcal{T}$ is the class label generator, and $\mathcal{L}_{ce}$ is the cross-entropy loss. 

Inspired by the next sentence prediction (NSP) in BERT \cite{devlin2018bert} (i.e., predicting if sentence B comes after sentence A), a few predictive self-supervised sequential recommendation models propose to predict the relation between two sequences. \textbf{PTUM} \cite{WuWQLH020} splits a user behavior sequence into two non-overlapping subsequences and predicts if a candidate behavior is the future behavior based on the past behaviors. \textbf{SSI} \cite{YuanCSZD21} pre-trains a Transformer-based recommendation model with a pretext task that shuffles/replaces a portion of items in a given sequence and predicts if the modified sequence is in the original order/from the same user. In the graph scenario, the pseudo-relation is often built by random-walks. \textbf{CHEST} \cite{wang2021curriculum} proposes conducting pre-defined meta-path-based random walks on heterogeneous user-item graphs to connect user-item pairs. It pre-trains a Transformer-based model with meta-path type prediction as the pretext task, predicting if there exists a path instance of a specific meta-path between a user-item pair.

\subsubsection{Similarity Prediction}
The similarity prediction task can be formulated as a regression problem, where the pre-computed/learned continuous values serve as the targets that the model's output needs to approximate. We can refine Eq. (\ref{eq: predictive}) to provide this branch of methods with the formulation as: 
\begin{equation}
    f_{\theta^{*}}=\underset{f_{\theta}, g_{\phi_{s}}}{\arg \min } \left\|g_{\phi_{s}}\left(f_{\theta}(\mathcal{D})\right)-\mathcal{T}(\mathcal{D})\right\|^{2},
\end{equation}
where $\mathcal{T}$ is the label generator that yields the targets. It can be an encoder to learn the targets or a series of actions to pre-compute the targets.

\textbf{BUIR} \cite{lee2021bootstrapping} is a a representative predictive SSR method inspired by the vision model BYOL \cite{grill2020bootstrap}, relying on two asymmetric graph encoders (an online network and a target network) to supervise each other without negative sampling. These two encoders are trained to predict the item representation output of each other, achieving self-supervision by bootstrapping representations. Particularly, the online network is updated in an end-to-end fashion while the target encoder is updated by momentum-based moving average to slowly approximate the online encoder. \textbf{SelfCF} \cite{zhou2021selfcf} inherits the merits of BUIR and further simplifies it by only using one shared encoder. To obtain more supervision signals to learn discriminative representations, it perturbs the output of the shared encoder. Another very similar concurrent work \textbf{CLUE} \cite{cheng2021learning}, which is an instantiation of BYOL \cite{grill2020bootstrap} in sequential recommendation, also employs one shared encoder. The main pipeline difference between SELFCF and CLUE is that CLUE augments the input and SELF augments the output representations.    

Similarity prediction is also used to address selection bias in recommender systems \cite{chen2020bias}. In \textbf{RDC}, pivot and non-pivot users are defined, where non-pivot users get biased recommendations by rating items they strongly like or dislike. RDC corrects the biases by forcing the rating distribution features of non-pivot users to be similar to those of their pivot counterparts, using dynamically computed rating distribution features as self-supervision signals \cite{liuhc2021self}. Similar tasks are also used for user preference disentanglement \cite{ma2019learning}. In \textbf{MrTransformer} \cite{ma2021improving}, each user preference representation is separated into a common and unique part, and then swapped to recombine the user preference representations. The model is trained to make the recombined preference as similar as possible to the original preference representations \cite{ma2021improving}. In heterogeneous information graphs, the similarity prediction task can capture rich semantics. In \textbf{DUAL} \cite{tao2022predictive}, meta-path-based random walks connect user-item pairs and record the number of meta-path instances as a probability measure of interaction. A path regression task is assigned to predict the pre-computed probability and integrate path semantics into node representations to enhance recommendation.

\subsection{Pros and Cons}
Compared to contrastive and generative methods that mainly rely on static augmentation operators, predictive methods acquire samples and pseudo-labels in more dynamic and flexible ways. Samples are predicted based on evolving model parameters, which refines self-supervision signals and aligns them with the optimization objective, likely improving recommendation performance. However, caution is needed when using pre-augmented labels. Most current methods \textbf{obtain pseudo-labels using heuristics, without assessing their relevance to recommendation tasks}. User-item interactions and associated attributes/relations generate with rationales, so expert knowledge should be incorporated into pseudo-label collection, which increases the expense of developing predictive SSR.

	\section{Hybrid Methods}\label{sec:Hybrid}
Hybrid methods assemble multiple types of pretext tasks to leverage different self-supervision signals. We divide surveyed hybrid methods into two groups according to how their self-supervised tasks interplay, including \textbf{Collaborative} SSR and \textbf{Parallel} SSR. A summary of surveyed hybrid methods is presented in Table \ref{Table: hybrid methods}.

\begin{table*}[t]
    \caption{A summary of the surveyed self-supervised recommendation methods with hybrid pretext tasks.}
	\label{Table: hybrid methods}
    \scriptsize
    \renewcommand\arraystretch{1.0}
    \begin{center}
        \resizebox{\textwidth}{!}{
    \begin{tabular}{lccccc}
        
    \toprule
    \textbf{Method} & \textbf{Scenario}  & \textbf{Data Augmentation} &   \textbf{Hybrid Type}  & \textbf{Branch} & \textbf{Training Scheme} \\  \hline
    CCL \cite{bian2021contrastive}    & Sequential      & Item Masking & Generative + Contrastive & Collaborative& Pre-training\&Fine-tuning              \\  \hline
    SEPT \cite{yu2021kdd}    & Graph (Social)      & \makecell{Edge Dropout\\Predicted samples} & Predictive + Contrastive & Collaborative&Joint Learning           \\  \hline
    COTREC \cite{xia2021cotrec}    & Sequential      & Predicted samples & Predictive + Contrastive & Collaborative& Joint Learning              \\  \hline
    CHEST \cite{wang2021curriculum}    & Graph      & \makecell{Node/Edge Masking\\Path Removal/Insertion} & Generative + Contrastive & Parallel&Pre-training\&Fine-tuning              \\  \hline
    MPT \cite{hao2021multi}    & Graph      & Node Masking/Substitution/Deletion & Generative + Contrastive & Parallel& Pre-training\&Fine-tuning   \\  \hline
   
    SSL \cite{YuanCSZD21}    & Sequential      & Item Substitution/Reordering & Predictive + Contrastive & Parallel&Pre-training\&Fine-tuning        \\ \hline
    PTUM \cite{WuWQLH020}    & Sequential      & Item Masking & Generative + Predictive & Parallel & Pre-training\&Fine-tuning        \\ \hline
    UPRec \cite{xiao2021uprec} &Sequential    & Item Masking & Generative + Predictive & Parallel & Pre-training\&Fine-tuning         \\
    \hline
    ODRec \cite{xia2022device} &Sequential    & Feature Mixing & Generative + Predictive & Parallel & Pre-training\&Fine-tuning         \\
    \bottomrule
\end{tabular}
}
\end{center}
\end{table*}

\subsection{Collaborative Self-Supervision}
To get comprehensive self-supervised signals, multiple self-supervised tasks collaborate to enhance the contrastive task by generating more informative samples under this branch. 

\textbf{CCL} \cite{bian2021contrastive} proposes a pre-training strategy for a Transformer-based model by linking a generative pretext task with a contrastive pretext task for sequential recommendation. The generative task involves masked-item prediction, with the predicted probabilities used to augment the sequence for contrasting to the original sequence. The method uses a curriculum learning strategy to arrange the contrastive task from easy to difficult based on the augmented sequences' ability to restore users' attribute information. \textbf{SEPT} \cite{yu2021kdd} is the first SSR model to integrate SSL and tri-training \cite{zhou2005tri} through a sample-based predictive task for social recommendation. It builds three graph encoders over three views with different social semantics. The encoders can predict semantically similar samples for the other two encoders, which are then used as positive self-supervision signals in the contrastive task. The improved encoders recursively predict more informative samples. \textbf{COTREC} \cite{xia2021cotrec} follows the same framework as SEPT but reduces the number of encoders to two for session-based recommendation. The two encoders are built over two session-induced temporal graphs and iteratively predict samples to enhance each other through the contrastive task. To prevent the mode collapse problem, COTREC uses an adversarial approach to keep the two encoders distinct.

\subsection{Parallel Self-Supervision}
In this branch, there are no correlations between different self-supervised tasks and they work in parallel. 

As an example model, \textbf{CHEST} \cite{wang2021curriculum} uses curriculum learning and self-supervised learning to pre-train a Transformer-based recommendation model on heterogeneous information networks. CHEST conducts meta-path-based random walks to form interaction-specific subgraphs and predicts the masked node/edge in a subgraph with the rest in the generative task, which exploits local context information. The contrastive task learns subgraph-level semantics by pulling the original and augmented subgraphs closer, exploiting global correlations. \textbf{MPT} \cite{hao2021multi} extends PT-GNN \cite{hao2021pre} to enhance representation capacity for cold-start recommendation. It adds contrastive tasks to capture inter-correlations across different subgraphs and augment data for training the GNN and Transformer-based encoders using node deletion/substitution/masking. The contrastive and generative tasks are conducted in parallel with their own parameters and then merged and fine-tuned.

In addition to the combination of generative and contrastive tasks, \textbf{SSI} \cite{YuanCSZD21} puts a label-based predictive task and a contrastive task together to pre-train models for sequential recommendation. It imposes two consistency constraints through predictive learning: temporal consistency and persona consistency. The former requires predicting if the input sequence is in the original order or shuffled, while the latter differentiates between an input sequence from a particular user and a sequence with unrelated items replacing some items. Additionally, an item masking-based contrast is conducted between the masked item representation and the sequence representation. \textbf{PTUM} \cite{WuWQLH020} imitates BERT \cite{devlin2018bert} to conduct both masked-item-generation and next-item-prediction tasks. \textbf{UPRec} \cite{xiao2021uprec} connects BERT with heterogeneous user information and pre-trains its encoder with social relation and user attribute prediction tasks. Besides these models, \textbf{ODRec} \cite{xia2022device}, as an on-device model, recombines server-side and device-side embeddings for self-supervision and unifies a contrastive task and a predictive task into a knowledge distillation framework. 

%In addition, there also has been research that combines generative tasks and predictive tasks. \textbf{PTUM} \cite{WuWQLH020} follows BERT \cite{devlin2018bert} to conduct the masked item generation task and the next behaviors prediction task. 

\subsection{Pros and Cons}
Hybrid methods assemble multiple self-supervised tasks to achieve enhanced and comprehensive self-supervision. Particularly, collaborative methods, where the generative/predictive task serves the contrastive task by dynamically generating samples, offer significant advantages in training effectiveness over static counterparts \cite{yu2021kdd,xia2021cotrec,bengio2009curriculum}. However, hybrid methods are confronted with the problem of \textbf{coordinating multiple self-supervised tasks}. They often struggle to balance different self-supervised tasks, requiring a manual search for hyperparameters or costly domain knowledge. Besides, different self-supervised tasks may also interfere with each other, which may require more complex architectures with a large number of parameters such as multi-gate mixture-of-experts networks \cite{ma2018modeling} to separate task-shared and task-specific information. As a result, a hybrid SSR model comes at a higher training cost compared to models with only a single self-supervised task. 

	\section{SELFREC: A library for Self-Supervised Recommendation}\label{sec:evaluation}
\begin{figure}[t]
    \centering
    \captionsetup{justification=centering}
    \includegraphics[width=.48\textwidth]{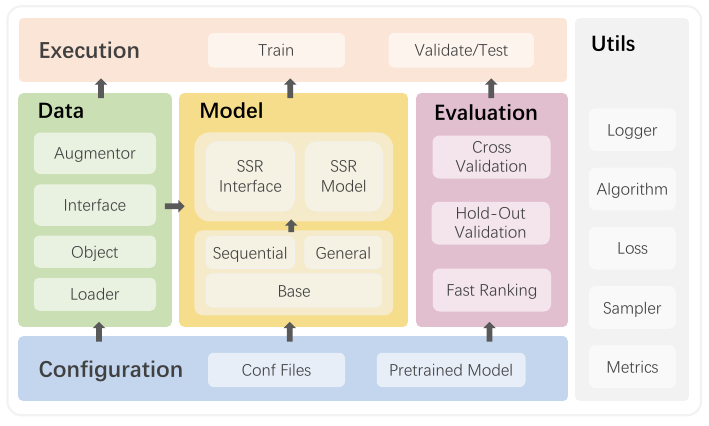}
    \caption{The architecture of SELFRec. }
    \label{fig:selfrec}				
\end{figure}

SSR is now enjoying a period of prosperity, with more and more SSR models mushrooming and claiming to be state-of-the-art. However, the empirical comparisons between different SSR models in the literature are often invalid due to inconsistent experimental settings, random selection of hyperparameters or even intentional modification of results of baselines. While some open-source repositories such as RecBole \cite{zhao2021recbole} and QRec \cite{yu2021kdd} have provided standard evaluation protocols, they are designed for universal purposes and their architectures are not ideal for implementing SSR models. For these reasons, we release an open-source library - \textbf{SELFRec}, which has an specialized architecture for SSR, shown in Fig. \ref{fig:selfrec}.

In SELFRec, we have incorporated several high-quality datasets that are widely used in the surveyed papers, such as Amazon-Book \cite{wujc2021self}, Yelp-2018 \cite{he2020lightgcn}, and Amazon-Beauty \cite{kang2018self}, for both general and sequential scenarios. We have integrated over 10 metrics, including ranking-based measures like MRR@$K$ and NDCG@$K$ and rating-based measures like MSE and RMSE. More than 20 SSR methods such as SGL \cite{wujc2021self}, CL4SRec \cite{xie2020contrastive}, and SimGCL \cite{yu2021graph} have been implemented in SELFRec for empirical comparison. Its important features are summarized as follows:
\begin{itemize}[leftmargin=*]    
    \item \textbf{Fast Execution}: SELFRec is compatible with Python 3.8+, Tensorflow 1.14+, and PyTorch 1.8+ and powered by GPUs. We also optimize the time-consuming item ranking procedure, drastically reducing ranking time to seconds.
    \item \textbf{Easy Expansion}: SELFRec provides simple and high-level interfaces, making it easy to add new SSR models in a plug-and-play fashion.
    \item \textbf{Highly Modularized}: SELFRec is divided into multiple discrete and independent modules. This design allows users to focus on the logic of their method and streamlining development.
    \item \textbf{SSR-Specific}: SELFRec provides specific modules for rapid development of data augmentations and self-supervised tasks.
\end{itemize}
Due to the limited space, we refer you to \url{https://github.com/Coder-Yu/SELFRec} for more information.

\section{Experimental Findings}\label{sec:experiment}
Considering the diversity of categories of data augmentation techniques and self-supervised learning methods, it is crucial to understand how to choose appropriate augmentation approaches and design self-supervised tasks for enhancing recommendation performance. Although this is not a paper centered on experiments and analysis, in this section we present some significant findings acquired through using SELFRec. These insights are believed to serve as guidelines for both researchers and practitioners.

We conduct a comparative analysis of prevalent data augmentation approaches and a set of representative SSR models from different categories. It is worth noting that the selection of backbone models can exert a more substantial influence on recommendation performance than other factors. For rigorous, meaningful and equitable comparisons, we select those graph-based models which employ LightGCN \cite{he2020lightgcn} as the backbone, while the chosen sequential models use Transformer \cite{vaswani2017attention} as the backbone due to their prevalence. With respect to the specific experimental configurations, we adopt the optimal hyperparameters reported in the original literature for the selected models and refined them through grid search. The general configurations like batch size, are aligned with those in \cite{wujc2021self,kang2018self}. We focus on the \textbf{Top-20} recommendation in both scenarios and rank all the items for an unbiased evaluation. The statistics of used datasets in SELFRec are shown in Table \ref{table:dataset}.

\begin{table}
    \begin{minipage}{\linewidth}
    \centering
    \caption{Dataset Statistics}
	\label{table:dataset}
	\scriptsize

		\begin{tabular}{c|cccc}
			\hline
			Dataset (Graph)&\#User & \#Item &  \#Feedback  & Density\\ \hline
			\hline
			Yelp2018 &31,668 &  38,048 & 1,561,406   & 0.13\%\\
			Amazon-Book & 52,463 & 91,599 & 2,984,108 & 0.06\%\\
            iFashion & 300,000  &81,614 & 1,607,813 & 0.007\%\\			
			\hline
		\end{tabular}

    \end{minipage}%
    \\
    \vspace{10px}

    \begin{minipage}{\linewidth}
        \scriptsize
		\centering
		\begin{tabular}{c|cccc}
			\hline
			Dataset (Sequence)&\#User & \#Item &  \#Feedback  & Density\\ \hline
			\hline
            Amazon-Beauty & 22,363 & 12,101 &  198,502 & 0.07\%\\		
			Amazon-Game &31,013  &  23,715 & 287,107   & 0.04\%\\
            Steam & 281,428 &  13,044 & 3,488,885   & 0.10\%\\
				
			\hline
		\end{tabular}
  \end{minipage}
    \end{table}

\subsection{Comparison of Data Augmentations in CL}
    \begin{table}[h]
	%\fontsize{9}{10}\selectfont
	\caption{Comparison of Graph Data Augmentations.}
	\small
	\label{table:graph-aug}
	\renewcommand\arraystretch{1.0}
    \resizebox{\columnwidth}{!}{
	\begin{tabular}{c|cc|cc|cc}
		\toprule
		\multirow{2}{*}{\textbf{Method}}&\multicolumn{2}{c}{\textbf{Yelp2018}}& \multicolumn{2}{c}{\textbf{Amazon-Book}} & \multicolumn{2}{c}{\textbf{iFashion}} \cr
		\cmidrule(lr){2-3}\cmidrule(lr){4-5}\cmidrule(lr){6-7}&\textbf{Recall} & \textbf{NDCG}  & \textbf{Recall} & \textbf{NDCG} & \textbf{Recall} & \textbf{NDCG}  \\ \hline
        LightGCN \cite{he2020lightgcn} &0.0639& 0.0525& 0.0410 & 0.0318 & 0.1053 & 0.0505	\\
		Node Dropout \cite{wujc2021self}  &0.0658& 0.0538& 0.0440 & 0.0346 & 0.1032 & 0.0498 \\
		Edge Dropout \cite{wujc2021self} &0.0675&0.0555& 0.0478 & 0.0379 & 0.1093 & 0.0531 \\	
		Random Walk \cite{wujc2021self} &0.0667&0.0547& 0.0457  & 0.0356  & 0.1095 & 0.0531  \\
		Feature Noise \cite{yu2021graph} &\underline{0.0688}& \underline{0.0570}& \underline{0.0488}  & \underline{0.0382}  & 0.1103 & 0.0532  \\
        Feature Dropout \cite{wujc2021self}&0.0659&0.0539& 0.0450  & 0.0353  & 0.1070 & 0.0509  \\
		Feature Mixing \cite{huang2021mixgcf} &0.0667&0.0549& 0.0464  & 0.0367 & \underline{0.1132} &\underline{0.0541}  \\		
		Layer Dropout \cite{yu2022xsimgcl} &\underline{0.0688} & 0.0566& 0.0485 & 0.0382 & 0.1071 & 0.0508   \\			
		\bottomrule
		\end{tabular}
}
\end{table}

\begin{table}[h]
	%\fontsize{9}{10}\selectfont
	\caption{Comparison of Sequence Data Augmentations.}
	\small
	\label{table:sequence-aug}
	\renewcommand\arraystretch{1.0}
    \resizebox{\columnwidth}{!}{
	\begin{tabular}{c|cc|cc|cc}
		\toprule
		\multirow{2}{*}{\textbf{Method}}&\multicolumn{2}{c}{\textbf{Amazon-Beauty}}& \multicolumn{2}{c}{\textbf{Amazon-Game}} & \multicolumn{2}{c}{\textbf{Steam}} \cr
		\cmidrule(lr){2-3}\cmidrule(lr){4-5}\cmidrule(lr){6-7}&\textbf{HR} & \textbf{NDCG}  & \textbf{HR} & \textbf{NDCG} & \textbf{HR} & \textbf{NDCG}  \\ \hline
        SASRec \cite{kang2018self}  &0.0980&0.0332& 0.1154 & 0.0391 & 0.1526 & 0.0602	\\
		Item Reordering \cite{xie2020contrastive} &0.0984&0.0343& 0.1153 & 0.0388 & 0.1557 & 0.0619	\\
		Item Masking \cite{xie2020contrastive} &0.0955&0.0329& 0.1130 & 0.0383 & 0.1568 & 0.0616 \\	
		Item Cropping \cite{xie2020contrastive} &0.0980&0.0337& 0.1135 & 0.0380  & 0.1566 & 0.0615  \\
		Item Substitute \cite{liu2021contrastive}  &0.0987&0.0344& 0.1165  & 0.0395  & 0.1568 & 0.0615  \\
        Feature Noise \cite{yu2021graph}	&0.1000 &0.0341& \underline{0.1178}  & \underline{0.0401}  & \underline{0.1588} & \underline{0.0636}  \\
		Feature Dropout \cite{Qiuwsdm22} &\underline{0.1020} & \underline{0.0348}& 0.1177  & 0.0396  & 0.1570 & 0.0626  \\		
		Layer Dropout \cite{liu2021modelaug} &0.0972 & 0.0330& 0.1158 & 0.0390 & 0.1570 & 0.0622   \\			
		\bottomrule
		\end{tabular}
}
\end{table}

In contrastive SSR, data augmentation approaches are rather diverse, including structure-level, feature-level, and model-level methods. As contrastive SSR is the dominant branch, in this part we investigate a number of most common data augmentations which can be used in a plug-and-play fashion. According to the results in Table \ref{table:graph-aug} and \ref{table:sequence-aug}, we can draw following conclusions:
\begin{itemize}[leftmargin=*] 
	\item Both scenarios show that augmentations at the feature level are highly effective. Specifically, adding noise to features results in the highest improvement on average.
	\item Augmentations at the structure level are less effective for sparser datasets and may even result in performance degradation in the sequential scenario. However, they can contribute positively to performance in denser datasets.
	\item The effectiveness of augmentations at the model level varies across different datasets. Some datasets show considerable improvement, while others show minimal improvement.
	\item Augmentations for sequential contrastive SSR are not as effective as its graph counterpart. One possible reason for this is the lack of clear semantics between item transitions, which may limit the potential of the Transformer structure.
\end{itemize}

\subsection{Comparison of SSR Models}
Different self-supervised tasks improves recommendation models from different perspective. To identify the most effective paradigm in SSR, we compared several popular SSR models in both graph and sequential scenarios. Our analysis of the results presented in Tables \ref{table:ssr-graph} and \ref{table:ssr-seq} yielded the following conclusions: (Note that we use (G, C, P) to indicate the category of the compared methods, referring to generative, contrastive, and predictive, respectively. We also manually constructed a LightGCN-based generative graph model (AdjRecons) adopting a structure generation task for fair comparison because no such published model exists.)
\begin{itemize}[leftmargin=*] 
	\item In the graph scenario, contrastive SSR methods demonstrate superior recommendation performance. Both SGL and SimGCL significantly improve LightGCN, but SimGCL shows greater superiority due to its more effective feature noise-based augmentations. In contrast, predictive SSR methods in this scenario are disappointing and even drastically lower the performance. We believe this is because predictive SSR requires more information, such as attributes, to create stronger and conducive self-supervision signals, but the used datasets do not provide attribute information. Generative SSR methods show performance that falls between contrastive SSR and predictive SSR, still achieving decent improvement.
	\item In the sequential scenario, contrastive and predictive self-supervised recommendation methods show similar performance improvements while the generative method BERT4Rec obtains disappointing results, even much lower than the results of SASRec. We believe this is because there is no fine-tuning in BERT4Rec. It is only pre-trained with the bidirectional masked-item prediction. However, there could be a gap between next-item prediction and bidirectional masked-item prediction.
	\item Compared to SSR methods in the graph scenario, there is still room for improvement in sequential SSR methods. Our findings suggest that many sequential SSR models are not as effective as reported in their original papers.
\end{itemize}

    \begin{table}[h]
	%\fontsize{9}{10}\selectfont
	\caption{Comparison of Graph SSR models}
	\small
	\label{table:ssr-graph}
	\renewcommand\arraystretch{1.0}
    \resizebox{\columnwidth}{!}{
	\begin{tabular}{c|cc|cc|cc}
		\toprule
		\multirow{2}{*}{\textbf{Method}}&\multicolumn{2}{c}{\textbf{Yelp2018}}& \multicolumn{2}{c}{\textbf{Amazon-Book}} & \multicolumn{2}{c}{\textbf{iFashion}} \cr
		\cmidrule(lr){2-3}\cmidrule(lr){4-5}\cmidrule(lr){6-7}&\textbf{Recall} & \textbf{NDCG}  & \textbf{Recall} & \textbf{NDCG} & \textbf{Recall} & \textbf{NDCG}  \\ \hline
		LightGCN \cite{he2020lightgcn} &0.0639& 0.0525& 0.0410 & 0.0318 & 0.1053 & 0.0505	\\
		AdjRecons (G)  &0.0665&0.0546& 0.0432 & 0.0336 & 0.1082 & 0.0523 \\	
		SGL (C) \cite{wujc2021self} &0.0675&0.0555& 0.0478  & 0.0379  & 0.1095 & 0.0531  \\
		SimGCL (C) \cite{yu2021graph} &\underline{0.0721}&\underline{0.0601}& \underline{0.0515}  & \underline{0.0404}  & \underline{0.1151} & \underline{0.0567}  \\
        BUIR (P) \cite{lee2021bootstrapping} &0.0487&0.0404& 0.0260  & 0.0209  & 0.0830 & 0.0384  \\
		SelfCF (P) \cite{zhou2021selfcf} &0.0525&0.0431& 0.0356  & 0.0283  & 0.0905 & 0.0437  \\
		\bottomrule
		\end{tabular}
}
\end{table}

    \begin{table}[h]
	%\fontsize{9}{10}\selectfont
	\caption{Comparison of Sequential SSR models.}
	\small
	\label{table:ssr-seq}
	\renewcommand\arraystretch{1.0}
    \resizebox{\columnwidth}{!}{
        \begin{tabular}{c|cc|cc|cc}
            \toprule
            \multirow{2}{*}{\textbf{Method}}&\multicolumn{2}{c}{\textbf{Amazon-Beauty}}& \multicolumn{2}{c}{\textbf{Amazon-Game}} & \multicolumn{2}{c}{\textbf{Steam}} \cr
            \cmidrule(lr){2-3}\cmidrule(lr){4-5}\cmidrule(lr){6-7}&\textbf{HR} & \textbf{NDCG}  & \textbf{HR} & \textbf{NDCG} & \textbf{HR} & \textbf{NDCG}  \\ \hline
            SASRec \cite{vaswani2017attention} &0.0980&0.0332& 0.1154 & 0.0391 & 0.1526 & 0.0602	\\
            BERT4Rec (G) \cite{sun2019bert4rec} &0.0718&0.0268& 0.0937 & 0.0400 & 0.1227 & 0.0477 \\	
            CL4SRec (C)\cite{xie2020contrastive} &0.0984&0.0343& 0.1153  & 0.0388  & 0.1568 & 0.0616  \\
            DuoRec (C) \cite{Qiuwsdm22}  &\underline{0.0996}&0.0342& \underline{0.1175}  & 0.0402  & \underline{0.1570} & \underline{0.0626}  \\
            ASRep (P) \cite{liu2021augmenting} &0.0988&0.0343& 0.1173  & \underline{0.0404}  & 0.1545 & 0.0609  \\
            BiCAT (P) \cite{jiang2021sequential} &0.0990&\underline{0.0345}& 0.1168  & 0.0398  & 0.1562 & 0.0620  \\		
            \bottomrule
            \end{tabular}
    }
    \end{table}
	
	\section{Discussion}\label{sec:discussion}
In this section, we identify the limitations of existing SSR methods and outline promising research directions that are worth exploring in the future.

\subsection{Theory for Augmentation Selection}
While data augmentation is essential for improving SSR performance, most current methods rely on heuristic approaches borrowed from other fields like CV, NLP, and graph learning. However, these approaches cannot be seamlessly transplanted to recommendation to deal with the user behavior data which is tightly coupled with the scenario and blended with noises and randomness. Besides, most methods augment data based on heuristics, and search the appropriate augmentations by the cumbersome trial-and-error. Although there have been some theories that try to demystify the visual view choices in contrastive learning \cite{Xiao0ED21,tian2020makes}, the principle for augmentation selection in recommendation is seldomly studied. A solid recommendation-specific theoretical foundation which can streamline the selection process and free people from the tedious trial-and-error work is therefore urgently needed. 

\subsection{Explainable Self-Supervised Recommendation}
Despite the promising results achieved by existing SSR models, the mechanisms behind their performance gains are not theoretically justified in most cases. These models are often considered black-boxes, with the primary goal being to achieve higher performance. However, components such as augmentations and self-supervised objectives lack reliable interpretability to demonstrate their effectiveness. Recent experiments in \cite{yu2021graph} have shown that some graph augmentations, which were previously thought to be informative, may even impair performance. Furthermore, it is unclear whether these models trade-off other properties, such as robustness, for their performance improvements. In order to create more reliable SSR models, it is crucial to understand what they have learned and how the model has been altered through self-supervised training.   

\subsection{Attacking and Defending Self-Supervised Recommendation Models}
Due to the open nature, recommender systems are vulnerable to the data poisoning attack which injects deliberately crafted user-item interaction data into the model training set to tamper with model parameters and manipulate recommendation results \cite{wang2022gray,zhang2021data}. The attack and the corresponding defense approaches for supervised recommender systems have been well-studied. However, it remains unknown if SSR models are robust to such attacks. We also notice that a few pioneering works have made attempts to attack the pre-trained encoders in vision and graph classification tasks \cite{saha2021backdoor,zhang2022unsupervised}. To ensure the robustness of SSR models, developing new attack strategies and corresponding defense mechanisms is urgent and significant.

\subsection{On-Device Self-Supervised Recommendation}
Modern recommender systems cater to millions of users through fully server-based operations, which come at a cost of a huge carbon footprint and raise privacy concerns. Decentralized recommender systems \cite{wang2021fast,wang2020next} have emerged as a solution by deploying on resource-constrained devices such as smartphones. However, on-device recommender systems are hindered by the highly compressed model size and limited labeled data. SSL can potentially address these problems, especially when combined with knowledge distillation techniques \cite{yu2021tiny,xia2022device,xia2023efficient} to compensate for accuracy degradation. Currently, on-device SSR remains less explored, and warrants further study.

\subsection{Towards General-Purpose Pre-Training}
In industry, recommender systems deal with multi-modal data and diverse scenarios. Deep recommendation models are trained across different modalities (e.g., video, image, and text) for various recommendation tasks \cite{huang2021once}. However, the training and tasks are often independent and require a large amount of computing resources. Given the relatedness of data across modalities, it is natural to explore general-purpose recommendation models pre-trained with multi-modal self-supervised learning on large-scale data. These models can adapt to multiple downstream recommendation tasks with cheap fine-tuning, making them particularly useful for scenarios with sparse training data. Although there have been efforts to develop general-purpose recommendation models \cite{shin2021one4all,zhang2020general,hou2022towards, hou2022learning}, they are mostly trained in a BERT-like fashion with similar architectures. It is worthwhile to investigate more efficient training strategies and effective model architectures.

\section{Conclusion}\label{sec:conclusion}
This survey paper provides a comprehensive review of the current state-of-the-art in self-supervised recommendation. It offers a taxonomy that categorizes existing SSR methods, an open-source library that facilitates empirical comparison, and some significant findings that shed light on the selection of self-supervised signals for enhancing recommendation. Finally, we outline future research directions to address the limitations of the current research.

	%\section{Acknowledgement}
	%This work was supported by ARC Discovery Project (Grant No.DP190101985, DP170103954).

	\ifCLASSOPTIONcaptionsoff
	\newpage
	\fi
	\bibliographystyle{IEEEtran}
	\bibliography{refs}
	
	\begin{IEEEbiography}[{\includegraphics[width=1in,height=1.25in,clip,keepaspectratio]{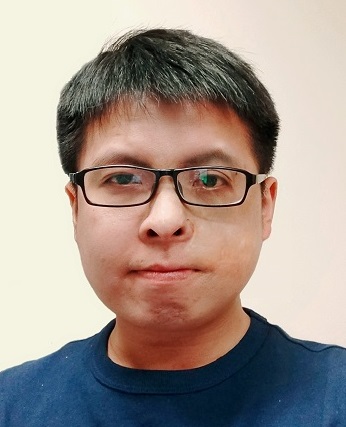}}]{Junliang Yu}
		completed his B.S. and M.S degrees at Chongqing University, and PhD degree at The University of Queensland. Currently, he is a postdoctoral research fellow at the School of Information Technology and Electrical Engineering, the University of Queensland. His research interests include recommender systems, data-centric AI, tiny machine learning, and self-supervised learning.
	\end{IEEEbiography}
	
	\begin{IEEEbiography}[{\includegraphics[width=1in,height=1.25in,clip,keepaspectratio]{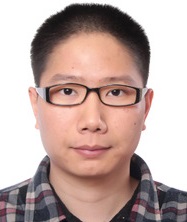}}]{Hongzhi Yin}
		 is an associate professor and ARC Future Fellow at the University of Queensland. He received his Ph.D. degree from Peking University, in 2014. His research interests include recommendation system, deep learning, social media mining, and federated learning. He is currently serving as Associate Editor/Guest Editor/Editorial Board for ACM Transactions on Information Systems (TOIS), ACM Transactions on Intelligent Systems and Technology (TIST), etc.
	\end{IEEEbiography}
	
	\begin{IEEEbiography}[{\includegraphics[width=1in,height=1.25in,clip,keepaspectratio]{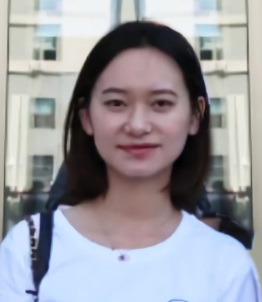}}]{Xin Xia}
		received her B.S. degree in Software Engineering from Jilin University, China. Currently, she is a third-year Ph.D. candidate at the School of Information Technology and Electrical Engineering, the University of Queensland. Her research interests include knowledge distillation, model compression, sequence modeling, and self-supervised learning.
	\end{IEEEbiography}

	\begin{IEEEbiography}[{\includegraphics[width=1in,height=1.25in,clip,keepaspectratio]{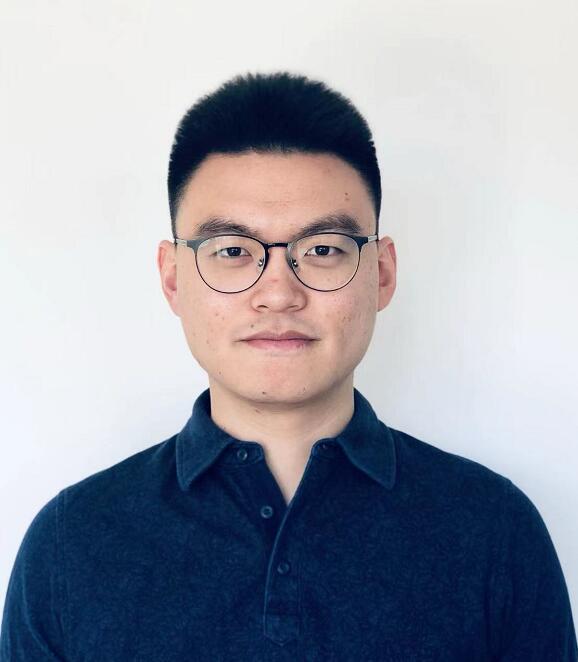}}]{Tong Chen}
		received his PhD degree in computer science from The University of Queensland in 2020. He is currently a Lecturer with the Data Science research group, School of Information Technology and Electrical Engineering, The University of Queensland. His research interests include data mining, recommender systems, user behavior modelling and predictive analytics.
	\end{IEEEbiography}

	\begin{IEEEbiography}[{\includegraphics[width=1in,height=1.25in,clip,keepaspectratio]{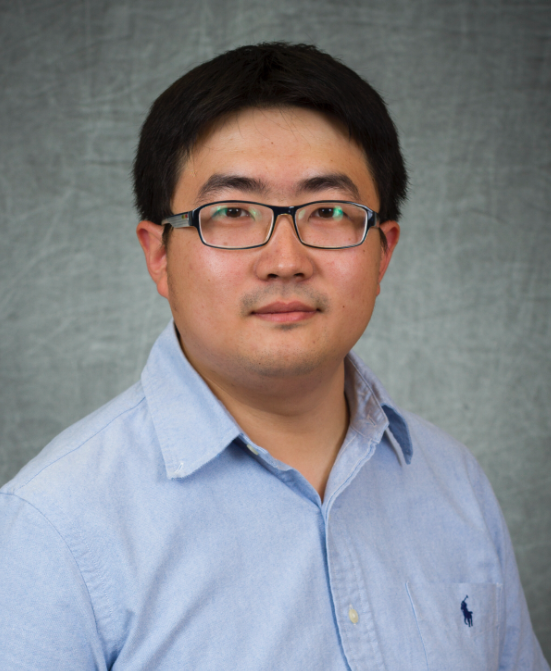}}]{Jundong Li}
		is an assistant professor at the University of Virginia. Prior to joining UVA, he received his Ph.D. degree in Computer Science at Arizona State University in 2019, M.Sc. degree in Computer Science at University of Alberta in 2014, and B.Eng. degree in Software Engineering at Zhejiang University in 2012. His research interests are broadly in data mining and machine learning, with a particular focus on feature learning, graph mining, and social computing.
	\end{IEEEbiography}

	\begin{IEEEbiography}[{\includegraphics[width=1in,height=1.25in,clip,keepaspectratio]{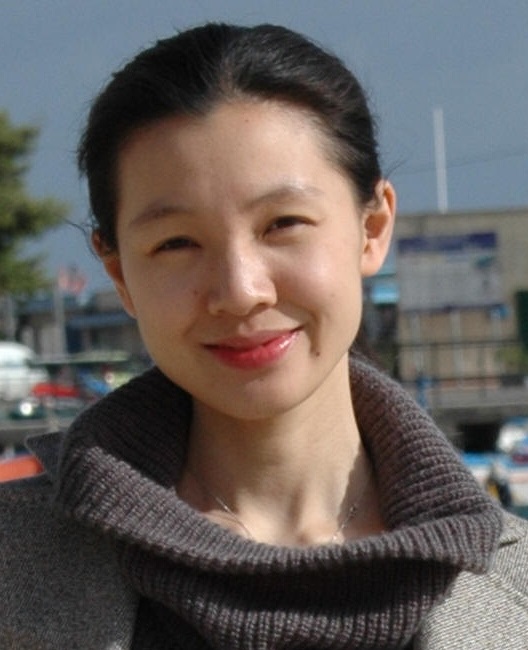}}]{Zi Huang}
		is a Professor and ARC Future Fellow in School of ITEE, The University of Queensland (UQ). She received her BSc degree from Department of Computer Science, Tsinghua University, China, and her PhD in Computer Science from UQ in 2001 and 2007 respectively. Dr. Huang's research interests mainly include multimedia indexing and search, social data analysis and knowledge discovery. She is currently an Associate Editor of The VLDB Journal, ACM Transactions on Information Systems (TOIS), etc.
	\end{IEEEbiography}

	%
	%% if you will not have a photo at all:
	%\begin{IEEEbiographynophoto}{John Doe}
	%Biography text here.
	%\end{IEEEbiographynophoto}
	%
	%% insert where needed to balance the two columns on the last page with
	%% biographies
	%%\newpage
	%
	%\begin{IEEEbiographynophoto}{Jane Doe}
	%Biography text here.
	%\end{IEEEbiographynophoto}
	
	% You can push biographies down or up by placing
	% a \vfill before or after them. The appropriate
	% use of \vfill depends on what kind of text is
	% on the last page and whether or not the columns
	% are being equalized.
	
	%\vfill
	
	% Can be used to pull up biographies so that the bottom of the last one
	% is flush with the other column.
	%\enlargethispage{-5in}

	% that's all folks
\end{document}